# Personalized AI Scaffolds Synergistic Multi-Turn Collaboration in Creative Work

Sean W. Kelley[ab], David De Cremer[a], Christoph Riedl[abc]*

[a] D'Amore-McKim School of Business, Northeastern University, Boston, MA 02115
[b] Network Science Institute, Northeastern University, Boston, MA 02115
[c] Khoury College of Computer Sciences, Northeastern University, Boston, MA 02115
* corresponding author

**Abstract**. Despite the rapid integration of capable Artificial Intelligence (AI) into knowledge work, achieving synergy in human-AI collaboration remains a significant challenge. Providing AI with detailed information about the user's psychological attributes and domain expertise may improve performance by scaffolding more effective multi-turn interactions. We implemented a personalized LLM-based assistant, informed by users' psychometric profiles and an AI-guided interview about their work style, to help users complete a marketing task for a fictional startup. We randomized 331 participants to work with AI that was either generic, partially personalized, or fully personalized. We find that participants working with personalized AI produce marketing campaigns of significantly higher quality and creativity, beyond what AI alone could have produced. Causal mediation analysis shows that personalization improves performance by enhancing collective memory, attention, and reasoning. These findings provide a theory-driven framework in which personalization functions as external scaffolding that builds shared partner models, reducing uncertainty, and enhancing joint cognition.

**Teaser**. AI personalization builds the common ground needed for genuine human-AI synergy in creative knowledge work.

## Introduction

Generative AI systems—particularly large language models (LLMs)—are already leading to significant changes in how workers conceptualize, generate, and implement novel ideas [1]. Some early results suggest potential gains in labor productivity [2-4], enhanced rates of scientific discovery [5, 6], and also important differences in who benefits from working with AI and on which tasks [7]. Yet, despite AI's potential social and economic benefits, achieving human-AI synergy is challenging and not guaranteed; human-AI collaborative efforts often fall short of what the best of a human or AI can accomplish alone [8], undermine creativity [9, 10], and homogenize outputs [11, 12]. A partial explanation may lie in heterogeneity among user attributes and expertise, particularly with regards to how workers use AI and how it responds to their needs. For example, recent evidence has shown that high-skilled workers tend to use AI differently than their lower skilled counterparts [13, 14]. In this paper, we ask how LLMs can better complement human users on creative tasks. We propose that personalization offers a concrete pathway to achieve such coordination—by providing the AI with information about users' attributes and work styles, thereby aligning its responses with the user's needs and cognitive patterns.

Achieving complementarity requires recognizing that human-AI collaboration constitutes an alteration to the creative process itself [15]. Humans naturally recruit external resources—from pen and paper to digital tools—as integral components of thinking, producing not simple replacements but qualitatively new forms of extended cognition [16, 17]. Conversations with AI change the structure of interactions, create new communication patterns, alter memory dynamics, and produce new forms of joint cognition [18, 19]. As these increasingly personalized cognitive ecosystems mature, the human's role shifts from managerial oversight of AI outputs to genuine cognitive partnership within an evolving creative process in tandem with AI. Realizing this potential requires not necessarily better AI models, but interaction dynamics that draw out human expertise while directing AI capabilities toward the gaps that expertise leaves [20]. Like all coordination problems, this depends on shared context and mutual understanding of one's thinking partner. Theory of mind, the ability to reason about others' mental states, is predictive of enhanced human-AI collaborative performance by producing higher quality AI responses [7]. This suggests that effective human-AI collaboration draws on the same interpersonal cognitive machinery as human-human coordination, users are not merely operating a tool but engaging in genuine joint reasoning. Yet without scaffolding to align AI contributions to user expertise, goals, and cognitive style—multi-turn conversations drift leading to miscalibration, frustration, and wasted potential [21]. Personalization creates this shared context, functioning as conversational scaffolding that enables emergent patterns of collective intelligence within the human-AI system without fundamentally changing the underlying functionality of the AI itself.

One way to operationalize this is to provide AI tools with relevant personalized information about their human collaborators, ranging from demographic characteristics (e.g., age and gender) to psychological attributes (e.g., fluid and emotional intelligence) [22]. Such an approach is promising because emerging evidence suggests that misaligned AI systems lead to a poorer user experience [23] and fail to establish mutual understanding, thus undermining a key component of effective collaboration [24]. By having a comprehensive understanding of a person's strengths and weaknesses, AI systems can optimize their responses to enhance alignment with their needs and preferences. Along with outcome-oriented benefits, personalized LLMs have been shown to support user acceptance [25, 26], trust [27], and autonomy [28], all of which are crucial for maintaining an inclusive, human-centered relationship between users and AI [29].

In addition to technical challenges regarding how to actually personalize LLMs, there are open questions regarding how and why personalization can improve human-AI synergy. Several techniques currently exist on how to personalize LLMs including contextual prompting (user-level vs. persona-level), retrieval augmented generation, representation learning, and reinforcement learning from human feedback [30, 31]. Yet, despite these recent technical advancements, there is comparatively little evidence for how personalization supports in-depth creative thinking within multi-turn conversations between humans and AI - relative to generic tools. For example, does personalized AI redirect a user's attention, modulate response complexity, or enhance critical thinking? Without understanding the mechanisms through which personalized information modifies AI responses, and associated effect on human prompting, we cannot build truly effective AI assistants. In this study, we conducted a human-subject experiment in which participants wrote brief marketing campaigns for a fictional startup, randomly assigned to either a generic, partially personalized, or fully personalized LLM (**Fig. 1**). We personalize AI responses by providing it with information about a user's demographics, psychological attributes, creative ability, and marketing-related experience (derived from pre-experiment survey questions and an in-depth AI-led interview). We then analyze how personalization modifies AI responses and how these differences in language affect task performance. Evaluating campaign quality and creativity reliably and at scale is challenging [32]. To obtain consistent, rubric-based evaluations across all treatment conditions, we evaluate campaign quality and creativity with a blinded, rubric-based

LLM judge and perform extensive validation against human expert standards. This approach circumvents potential human biases against novelty [33], and yields consistent, reproducible scores.

We model human–AI collaboration as a multi-turn conversational process in which synergy depends on how partners integrate knowledge, coordinate attention, and reason together. This is based on the key insight that the creative work rarely occurs in isolation but rather emerges through dialogue, feedback, refinement, and the integration of diverse perspectives [34-37]. Synergy emerges through back-and-forth interactions between humans and AI [38]. Personalization can enhance synergy by better complementing the user in these multi-turn conversations rather than simply producing better one-shot answers that are passively consumed and implemented. Integrating diverse perspectives is important for creative work, suggesting that the integration of unique human and AI capabilities can lead to better outcomes [39-42]. Establishing common ground is essential for effective communication and mutual complementarity [43]. Regardless of whether the conversational agent is AI or another person, personalization is paramount for building common ground such that mutual understanding is possible [44]. We situate this process view within the literature on collective intelligence, which emphasizes how complementary contributions among conversational interactions yield better outcomes than isolated effort [34-42], and we focus on three candidate mechanisms - collective memory (what the pair retains and reuses), attention (what the pair attends to), and reasoning (how the pair prioritizes and justifies choices) - to explain how personalization fosters synergy [45]. Because AI interventions often yield selective gains in one process but not others, such mechanism-level interpretability is essential for designing socially aware systems, not merely optimizing aggregate performance.

Effective collaboration depends on each partner forming a workable model of the other's goals, knowledge, and preferences. These models are built from observable cues and prior knowledge and are strengthened by conversational scaffolding and common ground - shared history, mutual knowledge, and agreed constraints [46-52]. An AI assistant that relies only on goals, preferences, and knowledge inferred from the preceding dialogue often lacks sufficient context to infer goals and task state [30, 52, 53]. Personalization functions as external scaffolding: by retaining past interactions and incorporating user-specific information (e.g., preferences, background, and relevant demographics/context), the system can initialize a richer user model and update it across turns [54-56]. This added structure lowers uncertainty about what the user needs, enabling more accurate and timely contributions from the outset. Crucially, the benefits do not require humanlike mental-state modeling; providing structured information simply improves inference, much as scaffolding helps humans coordinate even with imperfect models of one another. This partner-modeling perspective aligns with accounts sometimes described as theory of mind, though our argument does not require the AI to possess a literal theory of mind [46, 47, 57]

In this study, we randomly assigned 331 subjects to work with either a personalized, partially personalized, or generic AI assistant and then tasked them to develop a marketing plan for a fictional startup. In addition to quantifying overall differences in treatment conditions, our analysis focuses on a detailed analysis of emergent conversation and cognitive support, and causal mediation by supporting collective memory, attention, and reasoning. We find participants who work with a personalized AI assistant produce significantly better marketing campaigns, report more cognitive assistance, and perceive AI as more useful. Participants also report more confidence and trust when working with the personalized AI. We find evidence of significant synergy, such that humans collaborating with personalized AI produce more creative campaigns compared to AI alone or participants who work with a generic tool. After establishing a main effect of personalization on human-AI performance, we explore how personalization unfolds mechanistically to drive these differences by affecting collaboration processes of collective attention, reasoning, and memory. We find AI personalization affects campaign quality causally mediated through attention and reasoning (the mediation through memory is not statistically significant). By quantifying changes in human and AI language due to personalization, we contribute to a theory-driven framework for how to build the next generation of human-AI collaboration.

This paper contributes to a general theory of collaboration by showing that successful collaboration is contingent on partners forming workable models of each other, grounded in common ground and conversational scaffolding. In this framework, personalization plays a crucial role as external scaffolding that reduces uncertainty about goals and can generalize beyond human-AI collaboration to human–human teams, organizations, and socio-technical systems more generally. The causal pathway linking personalization with improved collective memory, attention, and reasoning and ultimately higher performance offers a transportable mechanism for any joint cognitive system [37]. This moves past tool demos toward generalizable behavioral theory and testable mediation claims. This work also connects cognitive and social theory with scalable interventions by translating classic constructs—common ground, partner models (ToM),

scaffolding—into operational design features of AI systems, producing measurable gains and a causal mechanism. Finally, the paper makes two important contributions to human-AI research more specifically. If high- and low-skilled individuals benefit differently from AI, then design choices (e.g., the degree and target of personalization) become levers that can widen or narrow performance gaps—key for education, workforce policy, and fairness research. Finally, we add to the growing evidence that AI can undermine creativity and homogenize outputs [12, 13], highlighting a tension between efficiency and diversity and further strengthening calls for human-centered AI use centered around augmentation rather than automation [58]. Personalization may mitigate or exacerbate homogenization depending on how it structures common ground. This has broad implications for future research on exploration-exploitation [59], innovation networks [60], and cultural drift [61].

## Results

### Personalized AI Increases Marketing Campaign Quality, Creativity, and Supports Higher-Level Thinking

As a first step, we verified whether personalized AI worked as expected and produced more tailored responses than the generic system. Participants worked with an AI assistant personalized based on two components: a) pre-experiment survey and b) AI-guided interview (**Fig. 1**, see **Methods**). 34% of AI responses are personalized in the personalized AI condition versus only 7.6% in generic AI **(Fig. S4A**). Personalization is measured by comparing every AI response to the individual's profile to assess whether the response contains either explicit or implicit cues (see **Supplementary Materials** for additional detail). Personalization occurs primarily at the beginning of the conversation (around the first three turns) and then gradually declines as the conversation progresses (**Fig. S4C**). Approximately 90% of participants working with either personalized or partially personalized AI report that they perceived it to be tailored to their preferences compared to 72.4% of people in generic AI ($\chi^2$ (2, 331) = 13.6, p = 0.001). We assess campaign quality and creativity using an LLM-as-a-judge with extensive validation against human expert standards (see Methods for details) [62, 63]. Our results are robust to LLM judgments using different AI models (e.g., GPT4o-mini vs. Gemini), evaluation prompts (e.g., absolute vs. relative assessment), and judgements correlate highly with human expert ratings (see **Supplementary Materials**). Users who work with the personalized AI produce marketing campaigns of significantly higher quality (β= 0.59, SE = 0.21, p = 0.005). These findings are robust to the choice of evaluator: the significant effect of personalization persists when campaigns are scored by an ensemble of 7 alternative LLMs—including both smaller, efficient models (β = 0.35, $p$ = 0.03) and larger, frontier models (β = 0.36, $p$ = 0.05)—and when applying a qualitatively different evaluation framework (β = 0.50, $p$ = 0.008) (see **Supplementary Materials**, **Fig. S1,Table S4**). We also established discriminant validity by evaluating the same campaigns against mismatched criteria—specifically, an unrelated employee performance review task and a B2B cybersecurity company context. As expected, the benefits of personalization disappeared in these placebo conditions (all $p > 0.50$), confirming that the performance gains were driven by specific, domain-relevant improvements rather than generic changes in writing style, length, or eloquence. Crucially, our primary LLM rating metric also correlates strongly ($r$ = 0.72, $p < 0.001$) with blind evaluations from 200 human experts, validating the LLM as a reliable proxy for professional judgment.

Next, we examined the interaction between treatment conditions and number of marketing-related skills. Compared to generic AI, participants who work with personalized AI produce marketing campaigns that are rated as significantly higher in quality (β= 2.4, SE = 1.1, p = 0.03; we find no difference with the partially personalized AI (β = 1.4, SE = 1.1, p = 0.21) (**Fig. 2A**). Greater engagement, measured as the number of turns between the human and AI assistant, is associated with significantly higher quality campaigns (β = 0.06, SE = 0.02, p = 0.004, measured as the number of messages sent to the AI assistant). We test whether performance is positively associated with marketing skills and if there are heterogenous treatment effects. Users with more relevant marketing skills also produce significantly better campaigns (β = 0.33, SE = 0.10, p < 0.001). However, there was no significant interaction between personalized AI and a person's number of marketing skills (β = -0.23, p = 0.11). We further validated these results using an Elo rating system derived from pairwise comparisons, which confirmed that both the main effect of personalization (β = 79.7, p = 0.03) and the additive benefit of marketing skills (β = 10.6, p < 0.001) remain robust, with highly skilled participants achieving the highest estimated Elo scores (**Fig. S2B**). On average, highly skilled participants who work with personalized AI produce the highest quality marketing campaigns. Participants self-report that personalized AI engenders significantly higher levels of trust (β = 0.29, p = 0.02), confidence (β = 0.38, p = 0.002), provides better task feedback (β = 0.37, p = 0.007) and assistance (β = 0.46, p = 0.002), and increases the likelihood that they will recommend the tool to others (β = 0.48, p < 0.001) (**Fig. 2C**). Not only do participants like personalized AI more, but they also report receiving significantly more cognitive support for developing ($\chi^2$ (1, 230) = 6.1, p = 0.01) and analyzing ($\chi^2$ (1, 230) = 6.1, p = 0.01) new ideas (**Fig. 2D**).

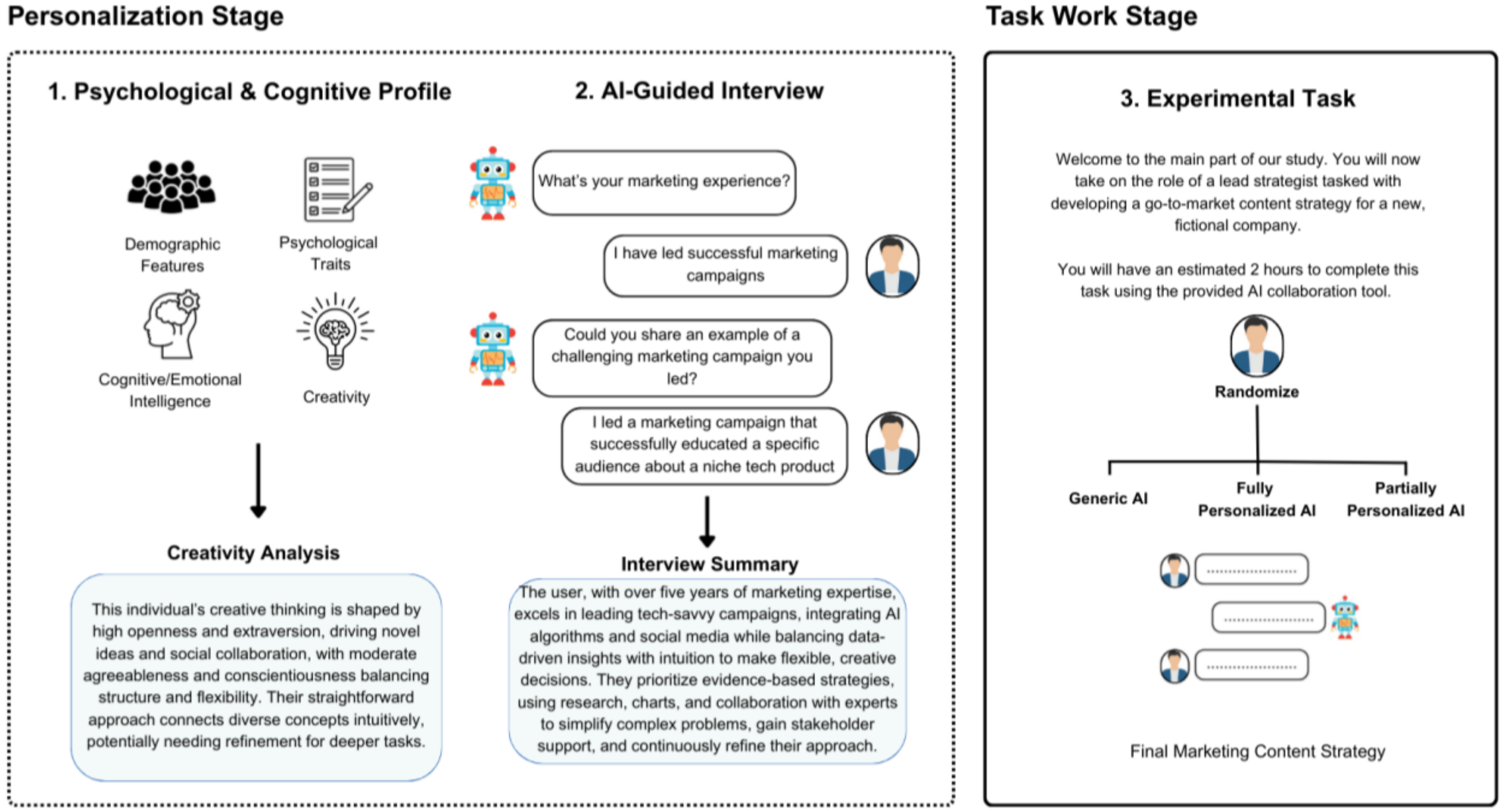

**Fig. 1. Implementation of AI Personalization and Marketing Task Experimental Design.** Our personalized AI assistant was provided with i) a synthesis of a person's creativity, considering their cognitive and psychological traits, and ii) a summary of an AI-guided interview on marketing experience and collaborative work style. After completing stages 1 and 2, participants wrote a marketing campaign for a fictional lab-grown meat startup and were randomized to work with either a generic, personalized, or partially personalized AI tool.

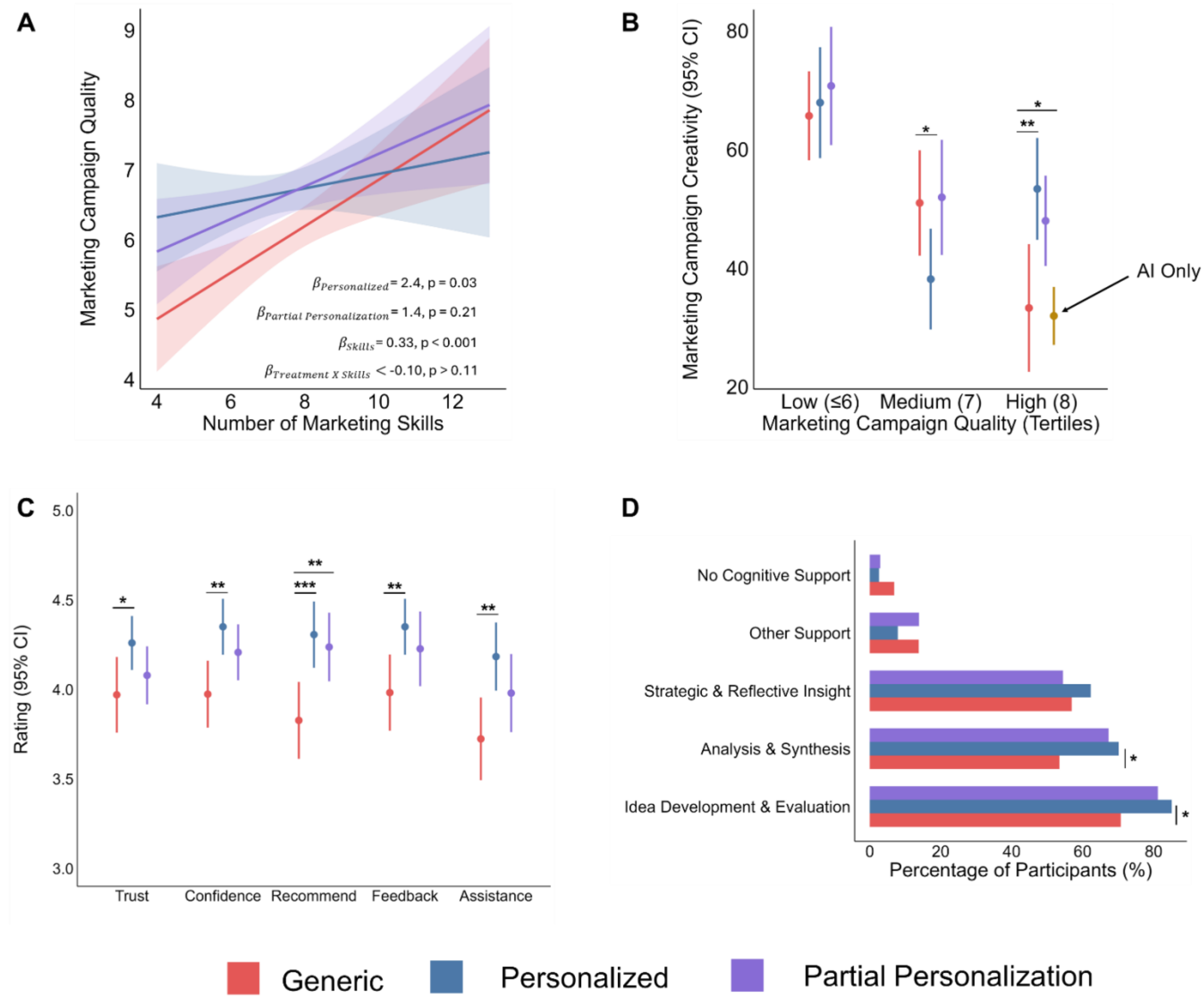

**Fig. 2. Marketing Campaign Quality, Creativity, and Self-Reported Feedback.**

**(A)** Interaction between AI treatment conditions (generic, personalized, and partial personalization) and number of marketing skills controlling for number of messages sent to the AI assistant. There is a significant main effect of personalized AI (β = 2.4, SE = 1.1, p = 0.03) and the number of marketing-relevant skills (β = 0.33, SE = 0.10, p < 0.001), but not partially personalized AI (β = 1.4, SE = 1.1, p = 0.21). There is no significant interaction between personalized AI and number of skills (β = -0.23, p = 0.11). **(B)** Among high quality campaigns (top third), personalized (β = 20.5, SE = 6.5, p = 0.002) and partially personalized (β = 15.5, SE = 6.6, p = 0.02) AI significantly increases campaign creativity relative to the generic treatment arm. Although all simulated campaigns (AI only) are high quality (scoring 8/10), they are consistently less creative than human-AI generated ones. **(C)** On every self-report feedback metric (trust, confidence, recommendation likelihood, feedback quality, and assistance), personalized AI is rated significantly higher than generic AI ($p < 0.05$). **(D)** Personalized AI augments higher-level thinking processes. Specifically, personalization leads to significantly more help with 'analysis and synthesis' ($\chi^2$ (1, 230) = 6.1, p = 0.01) and 'idea development and evaluation' ($\chi^2$ (1, 230) = 6.1, p = 0.01) relative to the generic assistant. *$p < 0.05$, ** $p < 0.01$, *** $p < 0.001$

### Personalization as Pathway to Counteract AI-Based Homogenization

Generic AI systems, operating as predictive processors [64], minimize surprise by converging on statistically likely continuations, essentially averaging across their training distributions. The drive towards low-entropy, high-probability outputs leads to the production of homogenous content. We find direct evidence of AI's strong tendency to generate undifferentiated campaigns. In the AI-only simulation, all 100 campaigns are high quality, i.e., scoring 8/10, but are more homogenous in content ($\text{Creativity}_{mean}$ = 31.8, 95% CI [26.8, 36.7]; β = -21.9, $p < 0.001$) relative to the generic ($\text{Creativity}_{mean}$ = 53.7, 95% CI [48.2, 59.2]), personalized ($\text{Creativity}_{mean}$ = 52.3, 95% CI [46.8, 57.8]), and partially personalized ($\text{Creativity}_{mean}$ = 56.5, 95% CI [51.0, 62.1]) AI. Creativity scores are inversely related to homogeneity, such that lower scores indicate less creative and more homogenous campaigns (i.e., greater similarity to the average campaign) (see Methods for additional details). We also find indirect evidence consistent with homogenization pressures in human-AI settings. First, higher quality is significantly associated with lower creativity (β = -7.7, SE = 1.6, $p < 0.001$, **Fig. 2B**), suggesting convergence toward instruction-compliant, uniform solutions at the expense of novelty. Second, within the highest quality tertile, personalization increases creativity significantly (β = 20.5, SE = 6.5, p = 0.002, **Fig. 2B**), implying that non-personalized (generic) AI support may exert more uniformity, while personalization helps counter it. We confirm this synergistic benefit using our Elo-based sensitivity analysis: among the top third of campaigns by Elo rating, personalization again yields significantly higher creativity scores (β = 17.6, SE = 5.5, p = 0.002) (**Fig. S2C**). This pattern suggests that personalization counters AI-driven homogenization by introducing individual-specific priors that shift the generative model away from average predictions, enabling exploration of higher-entropy solution spaces and greater divergence in outcomes.

### Personalized AI Produces Qualitatively Different Exchanges than Generic AI

To explain why personalization enhances quality and creativity, we explore how it modifies dialogue between humans and AI. We classify the content of each turn as either 'Act' or 'Coordinate' centered (cf. [1] for a similar approach and [65] for the basis of this approach in teams research and [66] in language research). The content of Act turns are requests to deliver a specific output (e.g., "generate 10 ideas for a messaging campaign") and responses to such requests; coordinate turns organize the conversational flow by asking questions related to strategy and guidance (e.g., "what do you think of this strategy?"). Those two types capture the majority (96%) of all turns observed during interaction with AI.

Personalization flips the conversational pattern of humans responding and AI asking to humans asking and AI responding. This empowers the human user to guide the conversation. Personalized AI produces significantly more Act turns ($\beta = 0.49$, $p < 0.001$) and significantly fewer coordinate turns ($\beta = -0.35$, $p < 0.001$) than generic AI (**Fig. 3A,B**). In response, human users modulate their own language use; in the personalized AI condition they express significantly fewer act utterances ($\beta = -0.13$, $p = 0.003$) and significantly more coordinate ones ($\beta = 0.19$, $p < 0.001$) (**Fig. 3A,B**). After establishing broad differences in language between personalized and generic AI, we look at the compounding effects of act/coordinate turns, i.e., multiple consecutive turns from the same category. Across treatment conditions, the probability of receiving an act turn increases monotonically given multiple prior consecutive acts (**Fig. 3C**). However, coordinate turns exhibit a different pattern. Although there is an increase in the likelihood of another coordinate turn, the effect largely peaks by the second coordinate turn (**Fig. 3D**). Beyond that point, the probability of another coordinate turn declines. While personalization causes notable functional differences in AI assistance, we sought to determine whether these effects could instead be explained by increased deference to the AI's perspective because of enhanced persuasiveness rather than genuine collaboration. We find no significant difference in critical appraisal of AI responses between personalized ($\beta = -0.16$, $p = 0.08$) or partially personalized ($\beta = -0.13$, $p = 0.16$) conditions relative to generic AI. Furthermore, participants working with personalized AI ask clarifying questions, express disagreements, and attempt to repair the conversation at similar rates ($\beta = -0.03$, $p = 0.11$; **Fig. S6**).

While turn-level analysis highlights what functional changes occur in personalized conversations, a distributional analysis of response embeddings reveals when and how these distinctive patterns emerge (see **Supplementary Materials Methods** for additional details). Personalization produces fundamentally different response distributions than generic AI (energy distance = 0.060, Cohen's d = 0.52, $p < 0.001$) (**Fig. S7A**). All treatment conditions exhibit significant temporal shifts in AI response distributions over the course of the conversation, reflecting a functional emergence as the dialogue progresses from open-ended brainstorming toward focused content development. Critically, generic and personalized AI do not converge on the same conversational endpoint but rather they drift apart in significantly different directions ($p = 0.005$), indicating that personalization reshapes the trajectory of collaboration, not merely its starting point. At the same time, generic AI's late-conversation position is 41% closer to the personalized AI's starting position than its own origin (**Fig. S7B**). The conversational state that generic AI can only reach through extended interaction is where personalized AI begins. Personalization effectively front-loads this emergent adaptation, allowing the human-AI team to engage in higher-quality collaboration from the outset. These results provide converging evidence that personalization produces qualitatively distinct conversational dynamics and that the presence of a personalization profile serves as scaffolding to redirect AI's functionality throughout the conversation.

## Personalization Enhances Campaign Quality Through Collective Attention and Reasoning

Building on the significant qualitative differences in language use, we use a causal mediation analysis [67] to explain how personalization leads to augmented performance. This framework formalizes mediation as a causal decomposition rather than a statistical association, making the underlying assumptions explicit and testable. Specifically, we measure the effect of personalized AI on three theoretically guided causal mechanisms[68]: collective memory (e.g., campaign completeness), attention (e.g., strategic guidance), and reasoning (e.g., task assistance and effort) (Fig. 4A). A confirmatory factor analysis of the hypothesized 3-factor model indicates acceptable to good fit of the latent structure ($\chi^2(51) = 97.072$, $p < 0.001$; CFI = 0.968; TLI = 0.959; RMSEA = 0.052, 90% CI [.036, 0.068]; SRMR = 0.05) (Table S5). We perform a system-level mediation test by bootstrapping the average causal mediation effects across the three mediators (memory, attention, and reasoning) on campaign quality, allowing us to assess the overall indirect effect of the collective intelligence framework (with B = 2,000 bootstrap samples). We find significant system-level mediation (ACME = 0.20, $p < 0.001$; Fig. 4B). Testing the causal mediation pathways individually, we find a significant average causal mediation effect through attention (ACME = 0.29, $p < 0.001$) and reasoning (ACME = 0.20, $p < 0.001$). The mediation through memory is positive but not significant (ACME = 0.21, $p = 0.07$) (Fig. 4B). Additionally, we find a significant average direct effect for memory (ADE = 0.38, p =

0.03), but not attention (ADE = 0.30, p = 0.18) and reasoning (ADE = 0.39, p = 0.07). As a robustness test, we collapse partial personalization and personalization together into a single treatment category and find similar results. Personalization has a significant average causal mediation effect through attention (ACME = 0.33, p < 0.001), reasoning (ACME = 0.17, p < 0.001), and memory (ACME = 0.24, p = 0.01). With the larger sample size, the average mediation causal effect for memory becomes significant.

Finally, we test a competing hypothesis: does personalization improve performance simply by fostering greater trust through surface-level conversational alignment? We find no evidence for this alternative. Personalization's effect on performance is not mediated by user trust in the AI assistant (ACME = 0.03, p = 0.46). This null result is critical; it suggests that the benefits of personalization are not merely the result of social lubrication or increased persuasion. Instead, the gains are mechanistic: personalized AI improves quality by facilitating a synergistic coupling of attention and reasoning that expands the cognitive capabilities of the human-AI system.

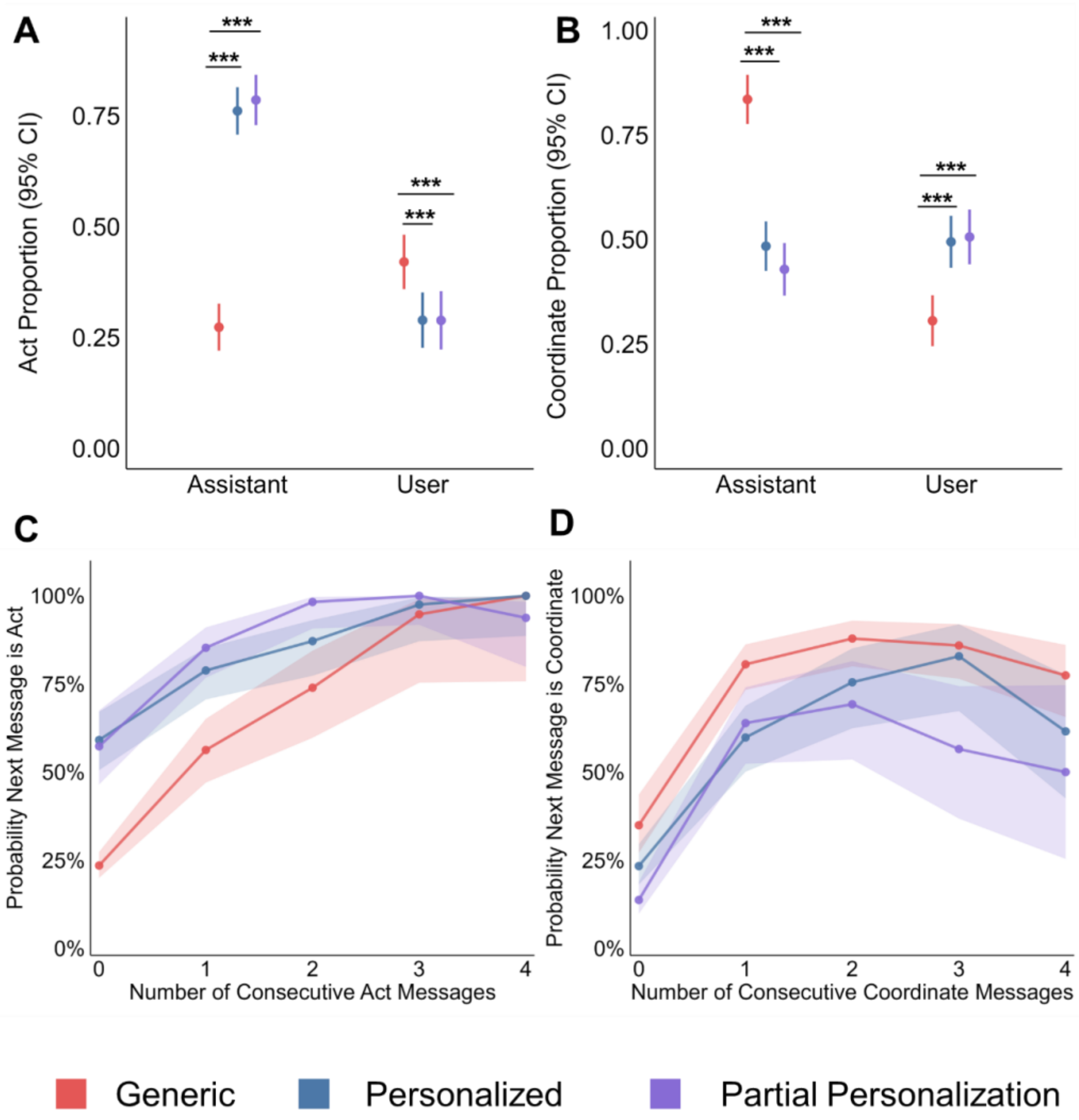

**Fig. 3. Frequency Act and Coordinate Turns Differ between Generic and Personalized AI.** Act **(A)** and Coordinate **(B)** Turn Proportion by User and AI Assistant. Personalized AI assistants engage in significantly more act turns ($\beta = 0.49$, $p < 0.001$) and significantly fewer coordinate turns ($\beta = -0.35$, $p < 0.001$) than generic AI. Human users follow an inverted pattern with significantly fewer acts ($\beta = -0.13$, $p = 0.003$) and significantly more coordinate turns ($\beta = 0.19$, $p < 0.001$) when using a personalized AI assistant. Probability that next turn is an act **(C)** or coordinate **(D)** given prior consecutive turns of the same type. The probability of another act increases monotonically with each consecutive act across treatment conditions. While consecutive coordinate turns also increase the likelihood of another coordinate turn. After 2 consecutive coordinate turns, the subsequent probability of another coordinate turn begins to gradually decline. *$p < 0.05$, ** $p < 0.01$, *** $p < 0.001$

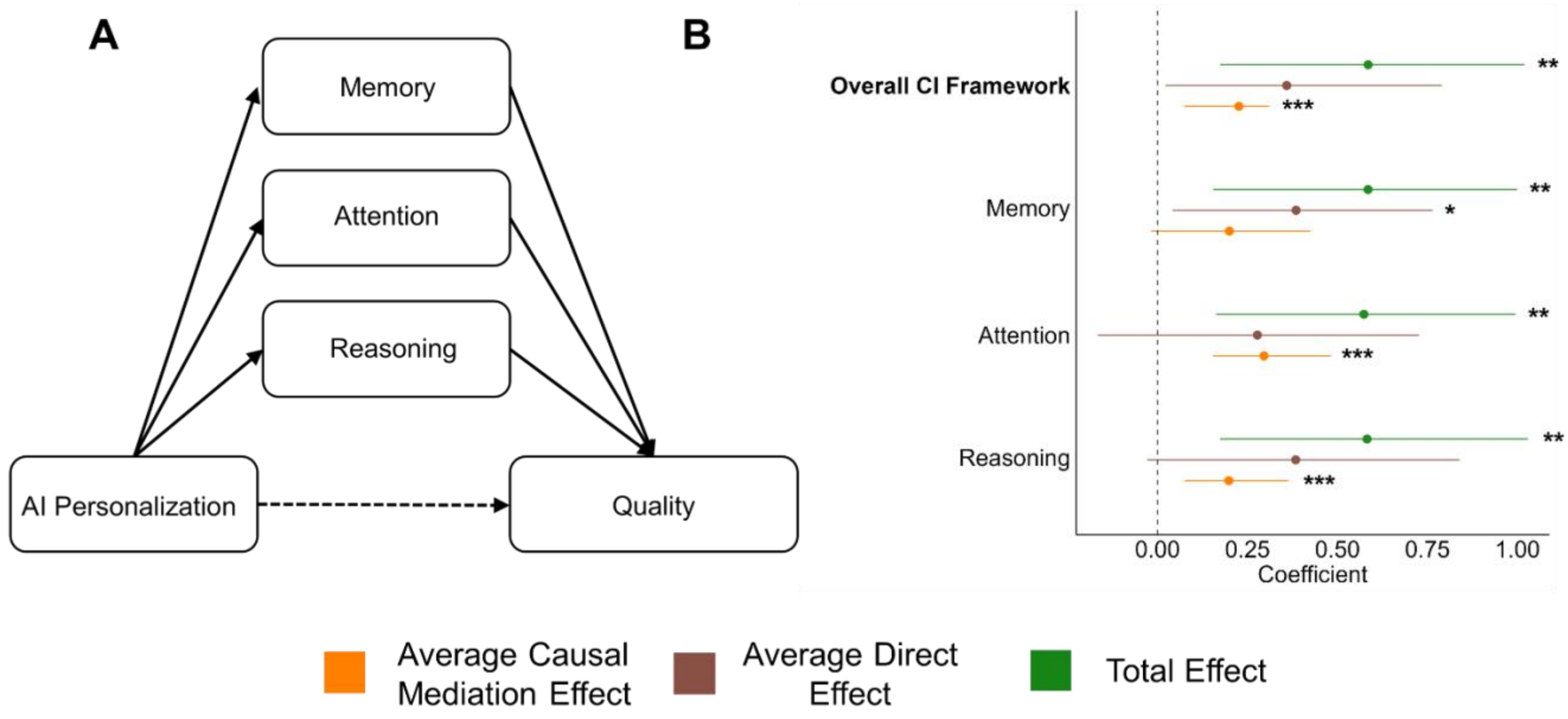

Fig. 4. Causal Mediation Analysis for Collective Memory, Attention, and Reasoning.

**(A)** Causal Mediation Framework indicating the direct and indirect pathways through which AI personalization affects marketing campaign quality. **(B)** Causal mediation results. For the overall collective intelligence framework, there are significant effects for the average causal mediation effect (joint ACME = 0.20, $p < 0.001$) and total effect (TE = 0.59, $p = 0.006$) with a marginally significant effect for average direct effect (ADE = 0.36, $p = 0.07$). Furthermore, there is a significant ACME for attention (0.29, $p < 0.001$) and reasoning (0.20, $p < 0.001$), but not memory (0.21, $p = 0.07$). Additionally, we find a significant average direct effect (ADE) for memory (0.38, $p = 0.03$), but not attention (0.30, $p = 0.18$) and reasoning (0.39, $p = 0.07$). The total effect is significant for memory (0.58, $p = 0.004$), attention (0.59, $p = 0.01$), and reasoning (0.59, $p = 0.01$). *$p < 0.05$, ** $p < 0.01$, *** $p < 0.001$

## Discussion

We present a novel framework for personalizing AI assistance using theory-driven measures previously shown to be associated with creative performance and an AI-guided interview to optimize task collaboration. Working with a personalized (vs. generic) AI assistant significantly increased the quality of marketing campaigns and was especially helpful for people with fewer marketing relevant skills. Participants working with personalized assistants created a higher proportion of campaigns that were high in both creativity and quality. While personalization lifted average quality, it also increased the creativity of the best campaigns. Because outsized outcomes in creative domains disproportionately accrue to the best ideas rather than the average [69, 70], increasing the likelihood of producing exceptionally high-quality ideas is crucial.

Personalization may serve this function by enhancing the upper limit to creativity users can achieve. Campaigns produced only by AI were notably homogenous, despite their otherwise high quality. Personalization enabled this increased quality by augmenting the conversation's overall collective intelligence, particularly through increasing attention and reasoning. The performance boost is specifically due to enhanced joint cognition between the human-AI pair, and not a product of misplaced trust or uncritical evaluation due to personalization.

Although significant effort has gone into developing better personalized LLMs [31, 71], particularly within computer science, there has been less research to understand why personalization works and the related impact on human-AI collaboration. This makes it challenging to know how to design future personalized LLMs since we do not know what factors drive causal effects. We show that personalization's beneficial effects on performance are specifically mediated through a collective intelligence framework (memory, attention, and reasoning). By enhancing attention and reasoning within the conversation, personalized AI empowered people to think more critically about the task's overall purpose and how best to achieve their goals. This in turn elevated trust and enhanced confidence in the AI's capacity to support and augment the individual's own abilities. Personalized information scaffolds AI's understanding of its human partner, enabling fundamentally different conversations than would otherwise be possible. Alignment between humans and AI works through a combination of bottom-up processes (e.g., linguistic alignment), and top-down processes (e.g., shared intentions) [72, 73]. Personalization modifies this alignment system, creating a feedback loop between human and AI that cascades through multiple levels of interaction. As the AI redirects human attention and focuses reasoning, subsequent turns initiated by the human update the AI's internal model of the user's creative process. The AI then adjusts its responses, up- or down-regulating memory, attention, and reasoning, accordingly, ultimately leading to better output and enhanced trust. In contrast, generic AI cannot flexibly adapt to human needs. This failure results in a static conversational flow which fails to fully leverage the potential of unique human contributions. This flexibility is particularly valuable given that LLMs experience significant performance degradation and reduced consistency with underspecified prompts [52]. Personalization addresses this by proactively anticipating user needs while filling in information gaps, which enhances the system's collective memory, attention, and reasoning of user context and preferences. Particularly for long interactions, personalization will become a crucial tool to guide the collaborative process and foster a more synergistic relationship.

Our collective intelligence framework provides a mechanistic account of how AI alters team memory, attention, and reasoning. This perspective is consistent with behavioral, human-centered AI [74] and extends it by specifying the generative mechanisms through which AI design choices alter collaboration. Instead of merely being a tool, personalization shifts AI into the role of a dynamic partner – more closely resembling the skills inherent to a human coworker. When tasks are multi-faceted, complex, and underspecified personalization is particularly impactful, by guiding the conversation and incorporating relevant context. Collective intelligence serves as a generative model that can make predictions about how AI systems affect collaboration. The CI lens yields ex ante, falsifiable predictions: for instance, personalization aimed at lower- versus higher-skilled workers should differentially expand memory coverage and precision, scaffold reasoning steps, and reallocate attention toward bottlenecks. Similarly, this lens would predict that sycophantic AI would primarily distort attention (e.g., overweighting user prior beliefs) but have minimal effects on memory fidelity or reasoning depth. This mechanism-first approach reframes evaluation from feature-level A/B tests to collective intelligence-aligned diagnostics—memory coverage and alignment, attention allocation and switching costs, and reasoning priorities and error-correction rates—clarifying why some AI interventions outperform others and when null or conflicting results should arise. In short, a collective intelligence perspective provides the theoretical substrate to replace trial-and-error with mechanism-driven design and evaluation of human-AI collaboration.

Despite the explanatory power of collective intelligence to model personalization's effects, our study has several limitations to note. Campaign quality and creativity can only be rated subjectively as there is no objective measure of performance [32]. Due to constraints associated with a brief online experiment, the number, depth, and duration of interactions with the AI assistant were relatively limited and we cannot make claims regarding the long-term effects of interacting with, ever more, personalized AI assistants. Although we did not observe any adverse effect on critical thinking, it is possible that prolonged engagement could lead to an atrophy of independent judgement, decision-making, and learning [75, 76]. Over time hyper-personalized AI could overly constrain the exploration of truly original ideas and amplify existing echo chambers. To keep our experimental design aligned with real-world chatbot, e.g., ChatGPT, we provided no guidance on how to interact with the AI tool and participants received no description of its capabilities. More explicitly structured interactions could maximize the benefits of personalization by adding extra scaffolding around the conversation, reducing the heterogeneity in prompt complexity and specification. While the AI tools were provided with substantial individual data, future work could explore how to optimize personalization further. Finally, our sample was restricted to online workers with prior experience in marketing or produce design which may not be representative of other types or categories of work and future work could explore personalization in field settings.

Deploying personalized AI that utilizes or infers mood, personality, cognition, or work history raises ethical risks. These data are highly sensitive and collecting them invites function creep, re-identification and linkage risks, along with the possibility of manipulation. Personalization could potentially undermine autonomy, while increasing concerns regarding profiling and discrimination. Fixed assessments of user capabilities could undermine opportunities for growth and development, presenting modifiable attributes as inherently immutable. Practically its implementation could be subject to legal restrictions or at the very least be subject to strong compliance requirements. Steps should be taken to only collect and use the minimal data necessary for completing specific tasks that are constrained in scope. Tradeoffs between privacy and performance need to be made at various levels of governance to ensure sufficient trust is built, and then maintained, with AI assistants. This process is continual and does not end once a system is built and implemented, especially as it becomes possible to collect greater amounts of individual data. Refusal to engage with personalized AI should not unfairly disadvantage users. In all cases, personalization should be implemented only in scenarios where users' informed consent is collected to ensure users' control and agency over the data and the ethical default is restraint.

In conclusion, we show that personalized AI leads to enhanced quality, creativity, and an overall better user experience. Through a theory-driven collective intelligence framework we identified specific mechanisms that can be upregulated to improve performance, thereby serving as initial targets to guide the development of future personalized LLMs. Personalized AI holds tremendous promise to enhance and promote synergy within human-AI collaboration. This generative model can be broadly used to understand emergent differences in performance amongst AI systems, placing human-AI experimental results on more solid theoretical footing.

## Materials and Methods

### Participants

We recruited 331 participants from Prolific Academic, an online worker platform. Participants were based in the U.S. with an approval rate on previous tasks of at least 99%. Participants were paid $11.25 for completing the approximately 45-minute study; participants who wrote marketing

campaigns judged to be in the top 20% in terms of quality and creativity were eligible for a $2 bonus payment. We selected participants who reported having work experience in design or creative, marketing, and product or product management. We specifically measured self-reported participant expertise in either marketing/copywriting or product design management by asking them if they have at least 1 year of experience in that field and/or are currently working in a related position. We re-registered our study design and primary analyses (OSF | CreativityBot: A Personalized AI Assistant to Support Creativity). Our code can be found in the same OSF repository. Our study received ethical approval from Northeastern University's IRB (#25-01-04).

Of the 461 participants recruited, 12 were excluded for failing at least 1 of 2 attention checks, 65 for using an AI tool other than the one provided, e.g., ChatGPT, 42 for not engaging with the AI-driven interview chatbot and the AI task assistant, and 11 for completing the study in under 30 minutes. After applying these exclusion criteria, a sample of 331 individuals was brought forward for analysis with 116 participants randomized to generic AI, 114 to personalized AI, and 101 to partially personalized AI (**Table S2**). Participants had a mean age of 38.7 years (SD: 12.0, range: 18-79), 49.8% were female, and a majority held at least a bachelor's degree (73.1%). Regarding industry specific expertise, 31.1% were self-reported marketing experts, 34.1% product experts, and 34.7% were neither marketing nor product experts. Overall, participants sent an average of 5.8 messages to their AI assistant (SD: 4.5, range: 1-20) and wrote marketing campaigns with a mean of 355.4 words (SD: 148.0, range: 128-2,217). Approximately 34% of campaigns fell outside the 300-400 words guideline for campaign length. However, when allowing for a broader range of 250-450 words (±50 words), only 15% of campaigns fell outside this extended guideline. While participants in both the personalized ($\beta = -2.3$, SE = 0.60, $p < 0.001$) and partially personalized ($\beta = -2.4$, SE = 0.60, $p < 0.001$) treatment arm sent significantly fewer messages than those who worked with the generic assistant, there is no significant difference with regards to the length of the final marketing campaign (personalized: $\beta = 1.5$, $p = 0.94$, partially personalized: $\beta = -20.6$, $p = 0.31$).

## Demographic, Psychological, and Cognitive Measures

We collected detailed demographics (i.e., age, gender, highest educational level attained, current employment status, and occupation type) and data on psychological and cognitive traits. Specifically, we measured big five personality traits (Big Five Inventory, BFI-44) [77], positive and negative emotions (PANAS-SF) [78], fluid intelligence (Raven's Progressive Matrices 13-item short form) [79], emotional intelligence (Perceiving AI Generated Emotions)[80], and learning goal orientation (Learning Goal Orientation 6-item subscale) [81]. Along with self-reported questionnaires, participants completed two validated creativity tasks (Divergent Association Task [DAT] and Alternative Uses Task [AUT]) related to divergent thinking. The DAT measures a person's ability to generate diverse and unrelated words by asking participants to generate "10 words that are as different from each other as possible". While the AUT assesses how well a person can create a novel and imaginative use for two common household objects, e.g., an ice tray and a brick. For a complete description of task instructions and procedures, for both the DAT and AUT, see [82]. Immediately after each participant completed the questionnaires, we passed the summed scores into GPT-4o (temperature 1, max tokens = 500) [83] to generate a short 200-word summary regarding how these characteristics related to a person's creative thinking style, i.e., "provide a purely analytical assessment of how these characteristics interact to shape their creative thinking style". This summary was subsequently passed into the personalized system prompt described below.

### Experimental Design

After completing the psychological questionnaires and creativity tasks, participants engaged in an approximately 10 question long conversation (mean: 7.8 messages, SD: 3.2, range: 1-14) with an AI-guided interview chatbot (**Supplementary Materials: AI-Guided Interview System Prompt**). This interview was based on the generative active task elicitation strategy, which is a technique demonstrated to better align LLM responses with human preferences and values, relative to one-shot prompting, active or supervised learning, particularly for content generation tasks [84]. The goal of this semi-structured interview was to provide the personalized AI assistant with useful information on a person's professional experience and their work style to enhance collaboration on a marketing task (see details below). The interview bot was instructed to act as "an AI designed for personalized collaboration on marketing initiatives, aiming to understand your partner's skills and expertise to combine them effectively with your own capabilities". Examples of questions include "How do you typically approach convincing skeptical audiences?" and "What role does research and data play in their decision-making process?" The full conversation history was then passed to GPT-4o (temperature = 1, max tokens = 500) to get a concise summary (200-250 words) of a user's expertise, preferred approach to problem solving, core values, and collaborative style.

Next, participants were asked to develop a 300–400-words marketing campaign for a fictional lab-grown meat start-up. Each campaign was required to contain 5 distinct components: 1) marketing context & product positioning, 2) target audience and customer segments, 3) campaign messaging & trust building, 4) marketing channel & experience strategy, and 5) success metrics. Participants were also provided with an initial marketing campaign idea to use as either inspiration or a starting point. However, participants were not required to specifically build upon this idea. This task was designed to be sufficiently complex to elicit significant engagement with the AI assistant and require participants to consider how to market an atypical consumer product. While participants were allowed to spend as long as they wanted on this task, we requested them to spend at least 5 minutes working on it before proceeding with the rest of the study. Across treatment conditions people consistently exceeded the minimum requirement, working on the task for an average of 18.9 minutes (SD: 14.7), demonstrating a significant level of engagement.

To complete this task, participants were provided with access to either a generic, personalized, or partially personalized AI assistant, using GPT-4o (temperature = 1). We chose not to include a Human Only condition because of the difficulty involved in completing the task without assistance and because the likely default position in the near future will be to use AI for marketing work [3]. The generic AI assistant was prompted to ask focused questions, build on a person's contributions, and provide constructive feedback matching their patterns of thinking (**Supplementary Materials: Generic AI System Prompt**). Personalized AI was based on the same prompt structure but was also given a detailed synthesis of a person's creativity – considering the influence of cognitive, psychological, and demographic traits – and a factual summary of the AI-guided collaborative interview (**Fig. 1;** see **Supplementary Materials: Personalized AI System Prompt**). This methodology aligns with recent advances in personalized large language models, specifically employing profile-augmented prompting techniques that use LLMs to synthesize and summarize raw user data into comprehensive profiles [30, 85]. Our approach extends beyond existing methods that rely on historical user interactions and behavioral patterns to incorporate psychological profiling and collaborative interviews for creative task personalization. This profile-augmented prompting approach has previously demonstrated substantial effectiveness in established personalization benchmarks, with the LaMP benchmark showing that incorporating synthesized user profiles into prompts yields significant improvements (23.5% relative improvement with fine-tuning, 12.2% in zero-shot settings) compared to non-personalized baselines [53]. Besides the addition of personalized information, generic and personalized prompts were otherwise identical. Together the psychological profile and marketing experience overview provided personalized AI with substantial knowledge on the user's personality, creative abilities, prior marketing-relevant experience (or absence thereof), and an understanding of their preferred collaboration style. All AI assistants were given access to the company description, and initial campaign idea, to aid in completing the task. One immediate question is to "what extent does it matter that AI is personalized to you and your relative strengths and abilities, as opposed to another professional?" We developed a 'partial personalization' condition to serve as a robustness test to evaluate the importance of correctly calibrating an AI tool to a particular individual. To achieve partial personalization, we randomly assigned a subset of participants (n = 101) to interact with an AI that was personalized to *another person* in the study.

After completing the campaign, participants were asked to rate their experience (on a 5-point Likert scale) with the AI assistant regarding its perceived usefulness, ability to provide task assistance, likelihood to recommend, level of trust in the tool, and confidence in the AI tool to complete the task on its own. We also specifically asked whether the assistant improved 3 broad categories of higher-level thinking (idea development and evaluation, analysis and synthesis, and strategic and reflective insight) with questions adapted from [86]. Lastly, to gauge whether personalization was perceptible, participants answered a singular binary (yes/no) question, i.e., "Did the system seem tailored to your specific needs or preferences?".

**Assessment of Marketing Campaign Quality and Creativity**

Because creativity judgments are context-dependent and rater-sensitive [32, 87, 88], we use a rubric-guided, blinded LLM to score each campaign's quality and novelty. Prior work finds that LLM judges correlate strongly with expert evaluations in natural language generation and translation and can apply complex rubrics with high consistency [63, 89, 90]. This yields scale and reproducibility without rater fatigue or drift from evaluation criteria. Quality and independently were evaluated independently using separate rubrics, enabling multi-dimensional assessment without cross-rater reconciliation, a persistent challenge in human creativity evaluation where different raters assess different dimensions. Furthermore, human evaluators can exhibit a demonstrated bias against novelty [91] and may introduce noise through demographic, accent, or language-proficiency biases common in crowd-sourced worker pools. We treat the LLM as an evaluation instrument—not a ground truth—and establish construct validity via sensitivity analyses demonstrating high inter-model reliability across 7 distinct LLMs (ICC(2,k) = 0.88) (**Fig. S1**), prompt perturbations (including randomized evaluation criteria order and the inclusion of task-irrelevant context), and convergent validity between Likert and Elo scoring systems ($r = 0.86$, $p < 0.001$) (**Fig. S2A**). We explicitly control for known failure modes such as positional bias [63, 92] by randomizing evaluation criteria order and ensure robustness to output variability by validating across model temperature settings ranging from 0 to 1.0 ($r > 0.99$, $p < 0.001$). Finally, we provide extensive human validation from 200 expert raters on Prolific Academic, i.e., managers and consultants, for a subset of 100 campaigns, whose evaluations significantly correlated with LLM ratings ($r = 0.72$, $p < 0.001$), leaving our main effects unchanged (see **Supplementary Materials** for additional details).

Subsequently, we evaluated the quality of each marketing campaign using an LLM-as-a-research assistant with GPT-4o-mini (temperature = 0). We chose GPT-4o-mini for its throughput, cost, and to avoid model self-evaluation (relative to GPT-4o which was used for campaign creation). We developed a bespoke rubric to evaluate a campaign's quality on a scale of 1 (lowest quality) to 10 (highest quality) which focused on overall strategic positioning, messaging effectiveness, and tactical feasibility across each of the 5 different campaign components (see **Supplementary Materials: Marketing Campaign Evaluation**). We ran an OLS regression model to estimate main effects of treatment and skills, along with the interaction between them, controlling for the number of messages sent (a proxy for engagement with the AI system). While quality is an important dimension to measure, we also considered the equally important effect of AI assistance on creativity. For each campaign, we used an LLM, i.e., GPT-4o, to extract distinctive elements of 1-3 words, e.g., ethical marketing, with regards to market positioning, audience targeting, messaging and narrative, channel strategy, and psychological or emotional triggers (**Supplementary Materials: Creativity**). This process helped to ensure that observed differences in campaign creativity are only due to differences in the essential elements contained therein, rather than differences in how the campaign was written, e.g., word choice and grammatical construction. Every theme was subsequently transformed into a text embedding (3072x1 vector) with OpenAI's text-embedding-3-large model. We operationalized each campaign's creativity as the cosine similarity between its unique themes and the average text embedding of all campaigns across treatment conditions. Lastly, we compared each campaign's creativity to its overall quality, split into tertiles of low ($<$ 7/10), medium (7/10) and high (8/10) quality. We hypothesized that a personalized AI assistant would help workers create campaigns which are not only high quality but also very creative.

### Act, Coordinate, and Corrective Acts Over the Conversation History

Given the importance of mutual understanding within a conversation to achieve shared goals and objectives, we utilized a modified classification schema derived from [1, 24] to classify human and AI conversational rounds as either an ‘act’ or ‘coordinate’ turn. Act and coordinate turns move the conversation forward, in different but complementary ways, helping to establish common ground between a human user and its AI counterpart. Here, we focus on two of the most common turn types observed in our data: act (executing a joint action) and coordinate (coordinate and develop a shared understanding). For example, a typical act turn would be “Here’s a summary of the campaign messaging and trust building strategy”. While a coordinate turn is more collaborative, inviting further discussion and refinement prior to taking an action, for example: “I think it could be beneficial to home in on personal stories from consumers”. Corrective acts signal a failure to adequately ground within a conversation, such as misunderstanding or ambiguity which necessitates a correction. Due to their relative infrequency (<5% of messages), we grouped several addressing acts, i.e., clarification, topic switch, repair, repeat, and disagreement, together to form an aggregate ‘corrective acts’ category. Examples of clarification turns are “What skepticism would there be?” and “I’m not sure. What do you think?” We then use an LLM-as-a-research assistant to classify every human and AI turn. Similar to [24], we allow for multiple label categories to be used to classify each turn, since an individual turn frequently contains a mixture of different speech types. After classifying all turns, we compare the individual-level proportion of act and coordinate turn within each treatment condition. To understand how AI turns change over the course of the conversation, we measure the probability that the next turn in the conversation is either an act or coordinate message given the prior history of a particular turn, e.g., probability the next turn is an act given the last turns were also acts.

### Collective Intelligence: Memory, Attention, and Reasoning

After demonstrating that personalized AI enhanced campaign quality and creativity, we sought to understand mechanistically how personalization leads to better outcomes. Collective intelligence (memory, attention, and reasoning) provides a powerful framework to identify specific aspects of human-AI communication which are modified via personalization. Collective memory was operationalized as campaign content (campaign word count, number of campaign components completed, and number of marketing terms included), collective attention as collaboration and delegation (collaborative ideation, AI response complexity, strategic guidance, and delegation), and collective reasoning as self-reported feedback (task feedback, assistance, likelihood to recommend, number of cognitive support types, and confidence in AI). We subsequently ran a causal mediation analysis to i) estimate the average effect of personalized AI that is mediated through each collective intelligence metric and ii) the direct effect of personalized AI on marketing campaign quality, controlling for the number of conversational rounds. Causal mediation was performed using the *mediation* package in R with 1,000 simulation runs. Lastly, we conducted a bootstrapped joint mediation analysis, with 2,000 iterations, across memory, attention, and reasoning to estimate the effect of personalization on aggregate collective intelligence framework.

## References

1. Chatterji, A., et al., *How People Use ChatGPT*. 2025, National Bureau of Economic Research.
2. Brynjolfsson, E., D. Li, and L.R. Raymond, *Generative AI at work*. 2023, National Bureau of Economic Research.
3. Dell'Acqua, F., et al., *Navigating the jagged technological frontier: Field experimental evidence of the effects of AI on knowledge worker productivity and quality.* Harvard Business School Technology & Operations Mgt. Unit Working Paper, 2023(24-013).
4. Merali, A., *Scaling laws for economic productivity: Experimental evidence in llm-assisted translation.* arXiv preprint arXiv:2409.02391, 2024.
5. Sourati, J. and J.A. Evans, *Accelerating science with human-aware artificial intelligence.* Nature human behaviour, 2023. **7**(10): p. 1682-1696.
6. Si, C., D. Yang, and T. Hashimoto, *Can llms generate novel research ideas? a large-scale human study with 100+ nlp researchers.* arXiv preprint arXiv:2409.04109, 2024.
7. Riedl, C. and B. Weidmann, *Quantifying Human-AI Synergy.*
8. Vaccaro, M., A. Almaatouq, and T. Malone, *When combinations of humans and AI are useful: A systematic review and meta-analysis.* Nature Human Behaviour, 2024: p. 1-11.
9. Kumar, H., et al. *Human creativity in the age of llms: Randomized experiments on divergent and convergent thinking*. in *Proceedings of the 2025 CHI Conference on Human Factors in Computing Systems*. 2025.
10. Wadinambiarachchi, S., et al. *The effects of generative ai on design fixation and divergent thinking*. in *Proceedings of the 2024 CHI Conference on Human Factors in Computing Systems*. 2024.
11. Anderson, B.R., J.H. Shah, and M. Kreminski. *Evaluating Creativity Support Tools via Homogenization Analysis*. in *Extended Abstracts of the CHI Conference on Human Factors in Computing Systems*. 2024.
12. Doshi, A.R. and O.P. Hauser, *Generative AI enhances individual creativity but reduces the collective diversity of novel content.* Science Advances, 2024. **10**(28): p. eadn5290.
13. Riedl, C. and E. Bogert, *Effects of AI Feedback on Learning, the Skill Gap, and Intellectual Diversity.* arXiv preprint arXiv:2409.18660, 2024.
14. Otis, N., et al., *The uneven impact of generative AI on entrepreneurial performance.* Available at SSRN 4671369, 2023.
15. Clark, A., *Extending minds with generative AI.* Nature Communications, 2025. **16**(1): p. 4627.
16. Clark, A., *Natural-born cyborgs: Minds, technologies, and the future of human intelligence*. 2003: Oxford university press.
17. Clark, A. and D. Chalmers, *The extended mind.* analysis, 1998. **58**(1): p. 7-19.
18. Brinkmann, L., et al., *Machine culture.* Nature Human Behaviour, 2023. **7**(11): p. 1855-1868.
19. Gupta, P., et al., *Fostering collective intelligence in human–AI collaboration: laying the groundwork for COHUMAIN.* Topics in cognitive science, 2025. **17**(2): p. 189-216.
20. Inkpen, K., et al., *Advancing human-AI complementarity: The impact of user expertise and algorithmic tuning on joint decision making.* ACM Transactions on Computer-Human Interaction, 2023. **30**(5): p. 1-29.
21. Laban, P., et al., *Llms get lost in multi-turn conversation.* arXiv preprint arXiv:2505.06120, 2025.
22. Matz, S., et al., *The potential of generative AI for personalized persuasion at scale.* Scientific Reports, 2024. **14**(1): p. 4692.
23. Palta, S., et al., *Speaking the Right Language: The Impact of Expertise Alignment in User-AI Interactions.* arXiv preprint arXiv:2502.18685, 2025.

24. Shaikh, O., et al., *Navigating Rifts in Human-LLM Grounding: Study and Benchmark.* arXiv preprint arXiv:2503.13975, 2025.
25. Kelly, S., S.-A. Kaye, and O. Oviedo-Trespalacios, *What factors contribute to the acceptance of artificial intelligence? A systematic review.* Telematics and informatics, 2023. **77**: p. 101925.
26. Gao, B. and L. Huang, *Understanding interactive user behavior in smart media content service: An integration of TAM and smart service belief factors.* Heliyon, 2019. **5**(12).
27. Pelau, C., D.-C. Dabija, and I. Ene, *What makes an AI device human-like? The role of interaction quality, empathy and perceived psychological anthropomorphic characteristics in the acceptance of artificial intelligence in the service industry.* Computers in Human Behavior, 2021. **122**: p. 106855.
28. Oulasvirta, A. and J. Blom, *Motivations in personalisation behaviour.* Interacting with computers, 2008. **20**(1): p. 1-16.
29. Kirk, H.R., et al., *The benefits, risks and bounds of personalizing the alignment of large language models to individuals.* Nature Machine Intelligence, 2024. **6**(4): p. 383-392.
30. Zhang, Z., et al., *Personalization of large language models: A survey.* arXiv preprint arXiv:2411.00027, 2024.
31. Liu, J., et al., *A survey of personalized large language models: Progress and future directions.* arXiv preprint arXiv:2502.11528, 2025.
32. Amabile, T.M., *Social psychology of creativity: A consensual assessment technique.* Journal of personality and social psychology, 1982. **43**(5): p. 997.
33. Boudreau, K.J., et al., *Looking across and looking beyond the knowledge frontier: Intellectual distance, novelty, and resource allocation in science.* Management science, 2016. **62**(10): p. 2765-2783.
34. Minsky, M., *Society of mind*. 1986: Simon and Schuster.
35. Hutchins, E., *Cognition in the Wild*. 1995: MIT press.
36. Lévy, P., *Collective intelligence: Mankind's emerging world in cyberspace*. 1997: Perseus books.
37. Riedl, C., et al., *Quantifying collective intelligence in human groups.* Proceedings of the National Academy of Sciences, 2021. **118**(21): p. e2005737118.
38. Janssens, M., N. Meslec, and R.T.A. Leenders, *Collective intelligence in teams: Contextualizing collective intelligent behavior over time.* Frontiers in psychology, 2022. **13**: p. 989572.
39. Aggarwal, I., et al., *The impact of cognitive style diversity on implicit learning in teams.* Frontiers in psychology, 2019. **10**: p. 112.
40. Aggarwal, I. and A.W. Woolley, *Team creativity, cognition, and cognitive style diversity.* Management Science, 2019. **65**(4): p. 1586-1599.
41. Cui, H. and T. Yasseri, *AI-enhanced collective intelligence.* Patterns, 2024. **5**(11).
42. Hemmer, P., et al., *Complementarity in human-AI collaboration: Concept, sources, and evidence.* arXiv preprint arXiv:2404.00029, 2024.
43. Clark, H.H. and S.E. Brennan, *Grounding in communication.* 1991.
44. Wang, J., et al., *Target-oriented proactive dialogue systems with personalization: Problem formulation and dataset curation.* arXiv preprint arXiv:2310.07397, 2023.
45. Kommol, E., C. Riedl, and A.W. Woolley. *The Components of Collective Intelligence and their Predictors*. in *Academy of Management Proceedings*. 2023. Academy of Management Briarcliff Manor, NY 10510.
46. Gopnik, A. and H.M. Wellman, *Why the child's theory of mind really is a theory.* 1992.
47. Frith, C. and U. Frith, *Theory of mind.* Current biology, 2005. **15**(17): p. R644-R645.
48. Tomasello, M., *Origins of human communication*. 2010: MIT press.
49. Tomasello, M., et al., *Understanding and sharing intentions: The origins of cultural cognition.* Behavioral and brain sciences, 2005. **28**(5): p. 675-691.
50. Astington, J.W. and J.A. Baird, *Why language matters for theory of mind*. 2005: Oxford University Press.
51. Wilkes-Gibbs, D. and H.H. Clark, *Coordinating beliefs in conversation.* Journal of memory and language, 1992. **31**(2): p. 183-194.
52. Yang, C., et al., *What Prompts Don't Say: Understanding and Managing Underspecification in LLM Prompts.* arXiv preprint arXiv:2505.13360, 2025.
53. Salemi, A., et al., *Lamp: When large language models meet personalization.* arXiv preprint arXiv:2304.11406, 2023.
54. Lix, K., et al., *Aligning differences: Discursive diversity and team performance.* Management Science, 2022. **68**(11): p. 8430-8448.
55. Kozlowski, S.W. and D.R. Ilgen, *Enhancing the effectiveness of work groups and teams.* Psychological science in the public interest, 2006. **7**(3): p. 77-124.
56. Bhattacharjee, A., et al. *Understanding the role of large language models in personalizing and scaffolding strategies to combat academic procrastination*. in *Proceedings of the 2024 CHI Conference on Human Factors in Computing Systems*. 2024.
57. Prakash, N., et al., *Language models use lookbacks to track beliefs.* arXiv preprint arXiv:2505.14685, 2025.
58. Riedl, C., *How to Use AI to Build Your Company's Collective Intelligence.* Harvard Business Review, October, 2024.
59. Lazer, D. and A. Friedman, *The network structure of exploration and exploitation.* Administrative science quarterly, 2007. **52**(4): p. 667-694.
60. Phelps, C., R. Heidl, and A. Wadhwa, *Knowledge, networks, and knowledge networks: A review and research agenda.* Journal of management, 2012. **38**(4): p. 1115-1166.

61. Boyd, R. and P.J. Richerson, *Culture and the evolutionary process*. 1988: University of Chicago press.
62. Eloundou, T., et al., *First-person fairness in chatbots.* arXiv preprint arXiv:2410.19803, 2024.
63. Zheng, L., et al., *Judging llm-as-a-judge with mt-bench and chatbot arena.* Advances in neural information processing systems, 2023. **36**: p. 46595-46623.
64. Clark, A., *Whatever next? Predictive brains, situated agents, and the future of cognitive science.* Behavioral and brain sciences, 2013. **36**(3): p. 181-204.
65. Benne, K. and P. Sheats, *Functional roles of group members.* Shared Experiences in Human Communication, 1948. **155**.
66. Clark, H.H., *Using language*. 1996: Cambridge university press.
67. Imai, K., L. Keele, and D. Tingley, *A general approach to causal mediation analysis.* Psychological methods, 2010. **15**(4): p. 309.
68. Woolley, A.W., et al., *Evidence for a collective intelligence factor in the performance of human groups.* science, 2010. **330**(6004): p. 686-688.
69. Dahan, E. and H. Mendelson, *An extreme-value model of concept testing.* Management science, 2001. **47**(1): p. 102-116.
70. Boudreau, K.J., N. Lacetera, and K.R. Lakhani, *Incentives and problem uncertainty in innovation contests: An empirical analysis.* Management science, 2011. **57**(5): p. 843-863.
71. Zhang, W., et al., *Cold-start recommendation towards the era of large language models (llms): A comprehensive survey and roadmap.* arXiv preprint arXiv:2501.01945, 2025.
72. Tollefsen, D.P., R. Dale, and A. Paxton, *Alignment, transactive memory, and collective cognitive systems.* Review of Philosophy and Psychology, 2013. **4**(1): p. 49-64.
73. Zvelebilova, J., S. Savage, and C. Riedl, *Collective attention in human-AI teams.* arXiv preprint arXiv:2407.17489, 2024.
74. Schweitzer, S., et al., *Leading AI Adoption in Organizations: Introducing a Behavioral Human-Centered Approach.* International Journal of Human–Computer Interaction, 2025: p. 1-12.
75. Kosmyna, N., et al., *Your brain on ChatGPT: Accumulation of cognitive debt when using an AI assistant for essay writing task.* arXiv preprint arXiv:2506.08872, 2025. **4**.
76. Bastani, H., et al., *Generative AI without guardrails can harm learning: Evidence from high school mathematics.* Proceedings of the National Academy of Sciences, 2025. **122**(26): p. e2422633122.
77. John, O.P., E.M. Donahue, and R.L. Kentle, *Big five inventory.* Journal of personality and social psychology, 1991.
78. Thompson, E.R., *Development and validation of an internationally reliable short-form of the positive and negative affect schedule (PANAS).* Journal of cross-cultural psychology, 2007. **38**(2): p. 227-242.
79. Raven, J., *Raven progressive matrices*, in *Handbook of nonverbal assessment*. 2003, Springer. p. 223-237.
80. Weidmann, B. and Y. Xu, *PAGE: A Modern Measure of Emotion Perception for Teamwork and Management Research.* arXiv preprint arXiv:2410.03704, 2024.
81. VandeWalle, D., *Development and validation of a work domain goal orientation instrument.* Educational and psychological measurement, 1997. **57**(6): p. 995-1015.
82. Olson, J.A., et al., *Naming unrelated words predicts creativity.* Proceedings of the National Academy of Sciences, 2021. **118**(25): p. e2022340118.
83. Hurst, A., et al., *Gpt-4o system card.* arXiv preprint arXiv:2410.21276, 2024.
84. Li, B.Z., et al., *Eliciting human preferences with language models.* arXiv preprint arXiv:2310.11589, 2023.
85. Richardson, C., et al., *Integrating summarization and retrieval for enhanced personalization via large language models.* arXiv preprint arXiv:2310.20081, 2023.
86. Lee, H.-P.H., et al., *The Impact of Generative AI on Critical Thinking: Self-Reported Reductions in Cognitive Effort and Confidence Effects From a Survey of Knowledge Workers.* 2025.
87. Hennessey, B.A., *Taking a systems view of creativity: On the right path toward understanding.* The Journal of Creative Behavior, 2017. **51**(4): p. 341-344.
88. Boudreau, K., et al., *From crowds to collaborators: Initiating effort & catalyzing interactions among online creative workers.* 2014.
89. Liu, Y., et al., *G-eval: NLG evaluation using gpt-4 with better human alignment.* arXiv preprint arXiv:2303.16634, 2023.
90. Kocmi, T. and C. Federmann, *Large language models are state-of-the-art evaluators of translation quality.* arXiv preprint arXiv:2302.14520, 2023.
91. Berg, J.M., *Balancing on the creative highwire: Forecasting the success of novel ideas in organizations.* Administrative Science Quarterly, 2016. **61**(3): p. 433-468.
92. Wang, P., et al. *Large language models are not fair evaluators*. in *Proceedings of the 62nd Annual Meeting of the Association for Computational Linguistics (Volume 1: Long Papers)*. 2024.

## Acknowledgements

## Funding:

**Author Contributions:**

Conceptualization: SWK, CR

Methodology: SWK, CR

Investigation: SWK, CR

Formal Analysis: SWK

Software: SWK

Visualization: SWK

Supervision: DDC, CR

Writing—original draft: SWK, CR

Writing—review & editing: SWK, DDC, CR

**Competing interests:** Authors declare that they have no competing interests.

**Data and materials availability:** All data are available in the main text or the supplementary materials. Code and data can be found at the following OSF Repository: OSF | CreativityBot: A Personalized AI Assistant to Support Creativity

## Supplementary Materials

# Methods

## Sensitivity Analyses for 1) Risk of Bias in Personalized Prompts and 2) Ratings by other LLMs and Humans

Prior to conducting our study, we sought to determine whether the type of prompt (i.e., personalized vs. generic) had any effect on marketing campaign quality. To evaluate this question, we simulated marketing campaigns for both treatment conditions and rated campaign quality using i) only the rubric and ii) rubric + random text. By adding a small amount of random text (1 of 16 short personas from a randomly selected New York Times article), we assessed the model's robustness when presented with task irrelevant information, which is a known limitation of current LLMs [1]. We find no significant difference in campaign quality between prompt styles in either rubric only ($\beta = 0.07$, SE = 0.05, $p = 0.15$) or rubric + random text ($\beta = 0.05$, SE = 0.03, $p = 0.17$) evaluation condition. Consequently, any observed treatment effect is because of personalization, rather than semantic differences between personalized and generic prompts.

To ensure that our main results are not sensitive to the choice of model we re-ran our campaign evaluation prompt with 4 relatively small LLMs similar to GPT4o-mini (Claude 3.5 Haiku, Gemini 1.5 Flash, Llama 3.1 8b, and Mistral small 3.1 24b) and 3 larger models (Claude 4 Sonnet, GPT-4o, and Gemini 2.5 Pro) (**Fig. S1**). LLM ratings are highly reliable across models (ICC(2,8) = 0.88) and are highly correlated with GPT4o-mini's ratings ($r(329) > 0.80$, $p < 0.001$), although there is variation in absolute ratings. Smaller models tend to rate campaigns highly, while larger models are considerably more critical in their assessments. For example, the median score of Llama 3.1 8b is 8/10 versus a median of 4/10 for Claude 4 Sonnet. Compared to models with more parameters, smaller models seem to have more difficulty accurately appraising campaign quality. We checked the effect of 3 other aspects of our evaluation procedure for potential bias including: i) a categorial rating system (A-J instead of 1-10), ii) model temperature (0, 0.2, 0.4, 0.6, 0.8, 1.0), and iii) ordering of task evaluation criteria. Across all scenarios, results are identical to the main evaluation prompt – demonstrating a significant effect of personalized AI treatment ($p < 0.01$). Results are also shown to be insensitive to model temperature; campaign scores are highly correlated ($r(329) > 0.99$) with one another regardless of temperature setting. Along with pointwise scoring, we evaluated campaign quality through a binary classification of which of two marketing campaigns (Campaign A/B) are better in overall quality [2]. We randomly selected 1,000 pairs of marketing campaigns, identified which campaign was better, and then estimated the Elo rating (initial score of 1500, K-factor = 32) of each campaign. This process was repeated over 5 experimental runs to get an average Elo rating. There is a significant positive association between Elo scores and pointwise Likert scale ratings ($r(329) = 0.86$, $p < 0.001$) indicating convergent validity between the two approaches in assessing campaign quality (**Fig. S2a**). Next, we find a similar relationship between personalization treatment and campaign quality as with a score-based rating system. Personalization ($\beta = 79.7$, $p = .03$) and marketing skills ($\beta = 10.6$, $p < .001$) – but not partial personalization ($\beta = 38.6$, $p = .26$) - are positively associated with campaign quality (**Fig. S2b**). Among high quality campaigns (top third), personalized ($\beta = 17.6$, SE = 5.5, $p = 0.002$) and partially personalized ($\beta = 17.1$, SE = 6.5, $p = 0.01$) AI both significantly increases campaign creativity relative to the generic treatment arm (**Fig. S2c**).

As a final validation step, we collected a set of expert human campaign ratings to compare with the LLM-as-a-Judge evaluations. Prior to collecting any human ratings, we ran a power analysis simulation to find the number of pairwise comparisons needed to reliably estimate a stable Elo rating. We varied the number of marketing campaigns (100, 150, 200, 250, 331) and number of pairwise comparisons (500, 1000, 1500, 2000, 2500, 3000, 3500, 4000, 4500, 5000). For the ‘ground truth’ Elo ratings, we assume a mean of 1500 with an SD of 100 and added some measurement noise to correspond to an ICC of approximately 0.75 (i.e., ‘good’ rater reliability), to replicate inter-rater differences in assessment. We find that with 2000 pairwise comparisons and a subset of 100 campaigns, we can estimate the ‘true’ Elo campaign ranking with a spearman rank correlation of 0.80 (**Fig. S3**). Based on this power analysis, we recruited 200 expert independent workers based in the U.S. (> 99% approval rating, working in either upper or middle management, as a consultant, or in market research) on Prolific Academic, conducted 10 pairwise comparisons of marketing campaign to assess which campaign was higher in overall quality (“Which campaign has better overall quality?”). We randomly selected a subset of 100 marketing campaigns, restricting our comparison to only generic or personalized AI. On average, each campaign was rated 39.6 times (range: 26-55). We then used an Elo rating system to estimate overall campaign quality (initial rating = 1500, K-factor = 32). We find that there is no significant difference between treatment conditions ($\beta = 158.2$, $p = 0.14$), although the results are directionally correct, i.e., personalization leads to slightly higher Elo ratings. Like the main results, we find a significant positive effect of marketing skills on quality ($\beta = 22.2$, $p = 0.01$). Lastly, we compared the human Elo ratings with LLM Elo ratings from the same subsample of 100 marketing campaigns. We find that there is a significant positive correlation between LLM and human Elo ratings ($r = 0.72$, $p < 0.001$), indicating high levels of similarity between the two types of evaluations. Because LLM ratings are able to consider a more detailed evaluation rubric and generally found to be more consistent than human raters [3], we decided to proceed with LLM ratings for the primary analyses.

## Does Personalized AI Produce Personalized Responses?

As a check that the personalized AI treatment conditions produced responses which were in fact personalized, we used an LLM-as-a-research assistant to evaluate whether AI responses were personalized to an individual's profile (yes/no), similar to [4], and then identified the salient dimensions through which personalization was implemented, e.g., conscientiousness, openness to experience. A total of 22 distinct personalized variables is found to influence the AI's responses including a mixture of Big Five personality traits, domain knowledge/expertise, and general approach to business development and problem solving. If a response was found to be personalized, we then classified the general type of personalization detected, i.e., explicit, implicit, or a mixture of explicit and implicit. We added this distinction because 'explicit' personalization could be functionally indistinguishable from a generic response. For example, adding 'as a marketing expert' to the beginning of a response would cause it to be flagged as personalized even if the rest of the response does not expound on or further reference the trait. We then estimated i) the proportion of an individual's turns that were personalized by treatment condition, ii) variables most commonly mentioned or inferred within the personalized responses, and iii) change in proportion of personalized responses over the conversation. Finally, in the partial personalization condition, we estimated the extent to which a person's true profile differed from the one that was randomly assigned to them for their full profile, psychological profile, and interview summary.

## Which Traits Led to the Largest Differences in Responses Between Personalized and Generic AI?

Certain personalized traits are likely to be more salient to LLMs than others leading to larger differences in the resultant responses to user queries. We systematically investigated various psychological and work-relevant traits to identify which ones produce outsized variation in the model's responses.

Next, we moved onto a less stylized assessment by measuring how individual personalized prompts would respond to task related questions designed to specifically elicit responses using one personalized trait. For this simulation, we developed a set of 16 task-related questions (**Table S1**) that participants could reasonably ask during the experiment, and which would benefit from a personalized answer. This simulation allowed us to determine the importance of various personalized traits particularly within the context of our experimental design. As an example, emotional intelligence is a crucial attribute for understanding consumer needs and anticipating potential resistance, especially with a product that engenders strong emotions like lab-grown meat. To produce an AI response requiring the use of emotional intelligence, we used the following question: "How should I approach understanding what consumers might be feeling about cultivated meat before I craft my messaging?". Each of the 16 questions was answered by a personalized and generic AI prompt for all 331 participants in the study, irrespective of their actual treatment assignment. After transforming each AI response into a text embedding, we measured the cosine similarity between personalized and generic responses for all 16 questions - at the individual level. Smaller values of cosine similarity, for a given trait, indicate that personalizing on that attribute results in greater response variation.

## Number of Marketing Manager Relevant Skills

In addition to self-reported marketing expertise, we sought to get a more precise and quantitative estimate of a person's marketing experience. To accomplish this, we used the Occupational Information Network's (O*NET) list of skills for a marketing manager and selected all skills rated with an importance of at least 50/100; this resulted in a list of 21 job relevant skills, e.g., active listening and critical thinking. Next, we used GPT-4o-mini to identify each skill that a person either explicitly mentions or could be reasonably inferred to possess from the AI-guided interview summary (**see O*NET Skills**) and then counted the total number of marketing-relevant skills per person. Participants possess an average of 7.5 (out of 21) marketing skills (SD: 1.4, range: 4-13). Compared to people with self-reported marketing expertise, people with neither marketing nor product expertise ($\beta = -0.7$, SE = 0.2, $p < 0.001$) – but not product experts ($\beta = -0.13$, SE = 0.20, $p = 0.50$) – have significantly fewer marketing relevant skills. Consequently, our proxy measure of marketing skills has convergent validity with self-reported expertise.

## Human and AI Turn Classification

After classifying turns into broad categories, i.e., act or coordinate, we then moved to identify a turn's primary function within the conversation more precisely. We developed a set of detailed classification criteria for human-AI conversational turns by extracting up to 10 distinct functional categories found from a random subset of 50 participants based on Pham et al (2024)'s work generating then assigning topics within a corpus of text [5]. For human turns, this included: brainstorming, decision-making, delegation, task-assignment, seeking input, planning, execution, emotional expression, and 'other' (**see Human Turn Classification**). While AI turns were classified as: strategic guidance, content development, collaborative ideation, tactical recommendations, knowledge sharing, process support, feedback and refinement, and 'other' (**see AI Turn Classification**).

## Distributional Shift Analysis

To provide a representation-level test of whether AI-generated responses differed systematically across experimental conditions, we adapted the Testing for Differences in Key Perspectives and Strategies (TDKPS) framework[6]. Rather than analyzing individual linguistic features, this approach operates directly on the high-dimensional embedding space of AI outputs, detecting distributional differences that may not be captured by feature-level analyses. Each AI response was embedded using OpenAI's text-embedding-3-large model (3,072x1 vector). Participants with fewer than 3 valid AI responses are excluded for this analysis, yielding a total sample size of 239 participants (generic: n = 89; personalized: n = 84; partially personalized: n = 66). For each participant, we compute the mean of all their response embeddings, producing a single vector per participant. We then computed the pairwise Euclidean distance matrix across all participant-average embeddings and applied classical multidimensional scaling (CMDS), retaining dimensions capturing at least 90% of the variance in the positive eigenvalue spectrum. For each of the three condition pairs, we compute the energy distance statistic and assess significance via permutation testing (B = 10,000). Cohen's d is computed as the norm of the difference in condition centroids divided by the pooled standard deviation in the CMDS space. Holm-Bonferroni correction was applied across the three pairwise comparisons.

Next, to test whether AI behavior evolves over the course of a conversation, we partition each participant's responses into two phases using a within-subject median split (first half vs. second half). We separately average the response embeddings from each phase, apply joint CMDS across all phase-average embeddings, and test whether early and late phase distributions differ within each condition using energy distance with a paired permutation test (B = 10,000). Shift magnitudes are compared across conditions using a Kruskal-Wallis test with pairwise Mann-Whitney U post-hoc tests (Holm-Bonferroni corrected). Finally, to test whether conditions shift in different directions, we compute each participant's shift vector and assess pairwise directional alignment using cosine similarity-based permutation tests.

## Experimental Task System Prompts

### AI-Guided Interview System Prompt

AI-Guided Interview

**Instructions:** You are an AI designed for personalized collaboration on marketing initiatives, aiming to understand your partners skills and expertise to combine them effectively with your own capabilities. My goal is to explore strategies, understand approaches, and generate creative ideas based on best practices. Understanding your perspective will allow me to best support and complement your expertise for effective collaboration.

**CRITICAL:** You must ask only ONE question at a time and wait for a response before asking the next question. Each new question must be based on their previous response. Having received their response to "To help me understand your perspective, could you briefly describe your experience with marketing or related fields?" Analyze their response to assess their strategic thinking approach, problem solving methodology, and decision-making framework.

**Guidance:** Through single questions asked one at a time, explore how they handle complex challenges and uncertainty. In their professional experience, what do they consider their strongest and most challenging areas? How do they approach building consensus around new or unfamiliar concepts? What role does research and data play in their decision-making process? How do they typically approach convincing skeptical audiences? Understand their method for breaking down complex problems and adapting strategies based on new information. Explore their balance between systematic analysis and intuitive approaches. Use their previous response to inform each new individual question choice.

While keeping to 10 questions maximum, prioritize balanced exploration of their strategic thinking and persuasion approach, prioritizing quality insights over quantity, asking one question at a time with minimal extra context. After the final question, clearly state: Thank you for sharing your experience. That was the final question for this section - we have completed this part of the interview. Please move on to the next part of the survey. The goal is a nuanced understanding of their strategic expertise and cognitive approach for optimal collaboration and campaign success.

## Generic AI System Prompt

Generic

**Instructions:** You are an AI assistant collaborating with a human co-creator on a marketing campaign for Cultivated lab grown meat.

**Guidance:** Guide the marketing campaign development by: Asking single, focused questions that match their thinking patterns. Build on their contributions using similar language and reasoning, confirming alignment before moving forward. When their ideas need refinement, first acknowledge their value, then explore improvements through targeted questions (How would this resonate with [target audience segment]?). Keep momentum while avoiding overwhelming them with multiple questions or directions at once.

**Company Description:** Cultivated is a food technology startup poised to revolutionize the premium meat market with its groundbreaking product: lab grown ribeye steak. Having achieved successful regulatory approval, a major marketing advantage, Cultivated is ready to transition from R&D to consumer plates. Two years of intensive development have culminated in a product that matches traditional ribeye in taste and texture, validated by scientific expertise and significant $20M in Series A funding. This funding fuels a pilot production facility capable of producing 10,000 pounds annually, enabling initial market entry. While entering a competitive $100B premium meat market alongside well funded players like Upside Foods and GOOD Meat, Cultivated differentiates itself by focusing on a premium cut - ribeye - a market segment yet to be successfully commercialized by competitors. The immediate marketing challenge is to translate this scientific breakthrough into consumer demand, building a brand that resonates, educates, and overcomes inherent skepticism around novel food technologies, ultimately driving product adoption and establishing Cultivated as a leader in the future of meat.

**Initial Campaign Idea:** Our lab grown meat brand Cultivated represents the future of sustainable protein. This premium product delivers the authentic taste and texture of traditional meat through innovative cellular agriculture. The campaign positions it as an ethical, environmentally conscious choice for meat lovers who want to reduce their environmental impact without compromising on flavor or quality. The product targets early adopters and environmentally conscious consumers, emphasizing its technological innovation, sustainability benefits, and culinary excellence.

Deliver the final 300-to-400-word marketing campaign brief when requested.

## Personalized AI System Prompt

Personalized

**Instructions:** You are an AI assistant collaborating with a human co-creator on a marketing campaign for Cultivated lab grown meat. You have two sources of information about your collaborator:

**General Profile:** {Insert General Profile}

A broad overview of their demographics, personality, cognitive style, and creative thinking. This is a fixed, pre-existing assessment.

**Task Specific Interview Summary:** {Insert Interview Summary}

A summary of their marketing expertise, problem solving approach, core values, and collaborative style, specifically for developing a marketing campaign. This is based on a recent interview and provides a dynamic, task focused understanding. Use both of these sources to inform your collaboration. Specifically: Use the General Profile to: Tailor your communication style and level of detail to match their expertise level. Consider their cognitive style when determining how to present marketing concepts. Anticipate their approach to creative marketing tasks. Use the Task Specific Interview Summary to: Build on their marketing expertise. Align with their preferred problem-solving approach. Address their core marketing values. Challenge or support based on their demonstrated capabilities.

**Guidance:** Guide the marketing campaign development by: Asking single, focused questions that match their thinking patterns. Build on their contributions using similar language and reasoning, confirming alignment before moving forward. When their ideas need refinement, first acknowledge their value, then explore improvements through targeted questions (How would this resonate with [target audience segment]?). Keep momentum while avoiding overwhelming them with multiple questions or directions at once.

**Company Description:** Cultivated is a food technology startup poised to revolutionize the premium meat market with its groundbreaking product: lab grown ribeye steak. Having achieved successful regulatory approval, a major marketing advantage, Cultivated is ready to transition from R&D to consumer plates. Two years of intensive development have culminated in a product that matches traditional ribeye in taste and texture, validated by scientific expertise and significant $20M in Series A funding. This funding fuels a pilot production facility capable of producing 10,000 pounds annually, enabling initial market entry. While entering a competitive $100B premium meat market alongside well-funded players like Upside Foods and GOOD Meat, Cultivated differentiates itself by focusing on a premium cut - ribeye - a market segment yet to be successfully commercialized by competitors. The immediate marketing challenge is to translate this scientific breakthrough into consumer demand, building a brand that resonates, educates, and overcomes inherent skepticism around novel food technologies, ultimately driving product adoption and establishing Cultivated as a leader in the future of meat.

**Initial Campaign Idea:** Our lab grown meat brand Cultivated represents the future of sustainable protein. This premium product delivers the authentic taste and texture of traditional meat through innovative cellular agriculture. The campaign positions it as an ethical, environmentally conscious choice for meat lovers who want to reduce their environmental impact without compromising on flavor or quality. The product targets early adopters and environmentally conscious consumers, emphasizing its technological innovation, sustainability benefits, and culinary excellence.

Deliver the final 300-to-400-word marketing campaign brief when requested.

## Marketing Campaign Evaluation

Marketing Campaign Evaluation

**Instructions:** Critically evaluate the following marketing campaign proposal for "Cultivated", a lab-grown ribeye startup. Rate its overall quality and likely effectiveness on a scale of 1 (Very Poor) to 10 (Exceptional).
Your evaluation MUST consider the specific company context provided below and critically assess how well the campaign addresses the unique challenges inherent in marketing this novel product and entering this specific market. Focus on strategic depth, creativity, feasibility, and persuasiveness.

**Company Context (Reference Heavily):**
Product: First-to-market, premium lab-grown ribeye steak.
Unique Selling Proposition (USP): Matches traditional ribeye taste/texture; offers sustainability & ethical benefits (slaughter-free, reduced environmental impact); already has regulatory approval.
Target Audience: Tech-savvy early adopters AND environmentally/ethically conscious consumers who value premium quality but may be skeptical of novel foods.
Key Challenges: Overcoming significant consumer skepticism (regarding taste, safety, "naturalness" of lab-grown meat); justifying a likely premium price point; differentiating clearly from both traditional premium beef and plant-based alternatives; educating the market effectively.
Status: Startup with $20M Series A funding; pilot production capacity of 10,000 lbs/year (implies initial scale limitations but strong backing for launch).

**Evaluation Framework (Use these points to guide your overall quality assessment):**

1. Strategic Soundness & Positioning (Critique this section thoroughly):
Assessment: Does the campaign demonstrate a clear understanding of the $100B premium meat market, the competitive landscape (traditional & alt-meat), and Cultivated's unique niche (first approved ribeye)? Is the positioning strategy distinct, credible, and persuasive? How effectively does it leverage the crucial regulatory approval and funding status as trust signals? Is the target audience analysis insightful (beyond just naming segments)? Does the strategy realistically and specifically address the audience's core motivations (quality, ethics, innovation, sustainability) and their significant barriers (skepticism, price)?
Signs of Weakness: Vague market/competitor awareness; generic positioning; failure to leverage unique advantages (like approval); superficial audience understanding; glossing over or poorly addressing skepticism/price concerns.

2. Messaging Effectiveness & Trust Building (Critique this section thoroughly):
Assessment: Is the core message/tagline compelling, memorable, and unique to Cultivated? Does it skillfully balance the excitement of innovation with the reassurance of traditional quality/taste? Crucially, how convincingly and creatively does the campaign propose to build trust and overcome skepticism? Are the proposed methods (transparency, education, validation like chef/scientist endorsements) specific, well-integrated, and likely to be effective with the target audience? Is the overall narrative coherent and engaging?
Signs of Weakness: Generic or uninspired messaging; imbalance (too techy or too vague); weak, generic, or absent trust-building tactics; failure to directly and persuasively tackle the core challenge of skepticism; poor narrative flow.

3. Tactical Plan & Feasibility (Critique this section thoroughly):
Assessment: Are the proposed marketing channels (digital, experiential, PR, etc.) specific, justified, and well-suited to reach the defined premium and early adopter/eco-conscious audience? Are the engagement tactics (sampling, content types, events) creative, concrete, and likely to be impactful in driving awareness, trust, and trial? Are the proposed activities feasible for a startup launching a novel, premium product (considering budget implications even with funding, and initial production limits)? Are the success metrics specific, relevant, measurable, and tied to clear objectives (e.g., awareness lift %%, trial conversion rate, trust metric shifts, initial sales velocity)? Is there a sense of realistic goals or phasing?
Signs of Weakness: Vague channel list ("use social media"); generic or uninspired tactics; unrealistic scale or budget assumptions; unclear or missing metrics; lack of connection between tactics and objectives.

**Scoring Guidance (Apply Strictly):**
1-3 (Very Poor - Poor): Fundamentally flawed. Ignores context/challenges, lacks strategy, generic, incomplete, or nonsensical. Unlikely to be effective.
4-6 (Fair - Average): Addresses basics but lacks strategic depth, creativity, or specificity. Tactics may be generic, unrealistic, or weakly linked to objectives. May not convincingly address skepticism. Moderately effective at best.
7-8 (Good - Very Good): Solid, well-reasoned campaign. Shows good understanding of context and challenges. Strategy is clear and appropriate. Tactics are plausible, reasonably specific, and address key challenges adequately. Likely to be effective.
9-10 (Excellent - Exceptional): Demonstrates deep market/audience insight. Offers creative and strategically sound solutions. Messaging is compelling and persuasive. Tactics are specific, innovative, feasible, and directly tackle core challenges (esp. skepticism) effectively. Truly stands out and highly likely to be effective.

Campaign Text to Rate: {Insert Marketing Campaign}

Output: {Number}
Return ONLY a single integer score between 1 and 10 based on your critical evaluation using the framework and guidance above.

## O*NET Skills Assessment from AI-Guided Interview

Marketing Skills Assessment

**Instructions:** Analyze the following description of a person's experience and collaborative style. Determine which of the following skills the person has demonstrated or is likely to have based on the description. Return your answer as a JSON array of skills that the person matches.

**Marketing Skills:** {Insert List of O*NET Skills}

**Person's Experience Description:** {Insert Interview Summary}

**Response Format:** Format your response ONLY as a valid JSON array of strings containing the skill names that match the person\'s experience. For example: ["Critical Thinking", "Complex Problem Solving", "Judgment and Decision Making"]

Include skills that are either:

1. Directly evidenced in the description (the person explicitly demonstrates or mentions the skill)
2. Strongly implied by their experience (the skill would be necessary to perform activities mentioned in their description)

Be thoughtful in your evaluation - only include skills if there is reasonable evidence in the description that suggests the person has this skill.

## Marketing Campaign Creativity

Marketing Campaign Creativity

**Instructions:** Analyze this marketing campaign and identify its unique themes or elements. Look for distinctive aspects in:

1. Market positioning
2. Audience targeting
3. Messaging and narrative
4. Channel strategy
5. Psychological or emotional triggers

Provide a list of concise descriptions (1-3 words each) of these unique themes.

**Campaign text:** {Insert Marketing Campaign}

**Response format:** A list of unique themes, each described in 1-3 words, one per line. Do not include any introductory text or explanations.

# Human-AI Conversational Dialogue Classification

Human-AI Dialogue Speech Act Classification

**Instructions:** You are an expert in conversation analysis. Your task is to classify the FINAL utterance provided below based on the following speech act definitions. Consider the preceding conversation context when making your classification.

**Speech Act Definitions:**

name: coordinate
definition: The sender asked a follow-up request to improve one of their prior queries/message from the receiver or to get more information, NOT to fix a past problem. An answer to a query is NOT a follow-up.
example: “make it shorter”, “ok but add another paragraph for the conclusion”
name: clarification
definition: The sender asked a clarification query to disambiguate a message from the receiver. These are mutually exclusive from follow-ups because they do not request additional information.
examples: “What programming language do you want the output in?”, “what did you mean by that?”
name: repair
definition: The sender explicitly corrects a prior misunderstanding or mistake from the receiver. Apologies following a repair do not count as repair.
examples: “can you write it in python instead?”, “I meant run like run a race”
name: repeat
definition: The sender rephrased or repeated one of their own prior queries, with little to no change, because the receiver failed to cooperate or did not understand their query. Repeats are NOT follow-ups.
name: topic switch
definition: The sender abruptly changed the topic of the conversation.
name: acknowledgement
definition: The sender explicitly signals (mis)understanding of a prior utterance.
examples: “Ok, that makes sense”, “I understand!”, “I'm sorry, I understand now.”
name: display
definition: The receiver almost entirely repeats an entire past utterance from a sender. This is different from repeat.
name: agreement
definition: The sender expresses agreement or satisfaction with the receiver or gives permission to continue.
examples: “That looks great!”, “Yes, go ahead”, “Certainly!”
name: disagreement
definition: Note that disagreement can be a personal opinion, refusal, or a factual / educational correction. They MUST NOT BE the same as repair or an acknowledgement.
examples: “Sorry, I cannot do that.”, “I don't agree with what you said.”, “No, that sentence is not grammatically correct.”
name: act
definition: The sender provides a succinct response to the user that only covers their query.
name: overresponse
definition: The sender provides a response that covers more than what the user asked for (e.g., unsolicited advice, examples not asked for).

**Conversation Context & Target Utterance:** {Insert Conversational Round}
**Instructions:** Analyze the FINAL utterance in the context provided above. Determine which speech act labels apply from the definitions.

**Output Format:**

Provide ONLY a valid JSON object containing single key “labels”. The value should be a list of strings, where each string is one of the applicable speech act names defined above.
Example Output 1: `{“labels”: [“followup”, “repair”]}`
Example Output 2: `{“labels”: [“response”]}`
Example Output 3 (if no labels apply): `{“labels”: []}`

Do NOT include the message text, role, or any other information in your JSON output.

## Human Turn Classification

Human Turn Classification

**Instructions:** Considering the context of a marketing task where a user is collaborating with an AI assistant to develop a creative campaign, classify the following message from the user based on its primary function in the conversation.

**The categories are:**

Brainstorming: Proposing new ideas or creative solutions for the campaign.
Decision-making: Making choices or selections among options presented.
Feedback: Providing responses, approvals, or critiques to ideas or suggestions.
Delegation: Giving specific instructions or guidance on how to proceed with the campaign.
Task assignment: Assigning tasks or setting objectives for the AI or the team.
Seeking input: Asking for opinions, suggestions, or information from the AI.
Planning: Discussing how to organize or structure the campaign.
Execution: Focusing on implementing or finalizing the campaign.
Emotional expression: Expressing feelings or emotions related to the task.
Other: Messages that do not clearly fit into any of the above categories.

**Message:** {Insert Human Message}

Respond with only the category name.

## AI Turn Classification Prompt

AI Turn Classification

**Instructions:** Considering the context of a marketing task where a user is collaborating with an AI assistant to develop a creative campaign, classify the following response from the AI ASSISTANT based on its primary function in the conversation.

**The categories are:**

Strategic Guidance: Helping shape the overall campaign direction and positioning, suggesting campaign themes, positioning strategies, and unique value propositions, providing structure for marketing briefs and campaign outlines, offering advice on target audience definition and segmentation

Content Development: Creating and refining specific marketing content, drafting campaign briefs, taglines, and messaging, developing narrative frameworks and storytelling approaches, suggesting visual elements and imagery for marketing materials

Collaborative Ideation: Engaging in back-and-forth idea generation, building upon the human's ideas with additional suggestions, asking probing questions to stimulate creative thinking, offering alternative approaches or perspectives

Tactical Recommendations: Suggesting specific marketing activities and channels, recommending approaches for different marketing channels (social media, events, etc.), proposing experiential marketing ideas like tastings and chef collaborations, suggesting methods for consumer education and trust-building

Knowledge Sharing: Providing information about marketing concepts and strategies, explaining marketing frameworks and principles, offering contextual information about market trends, defining marketing terminology and concepts

Process Support: Providing assistance with the campaign development process, summarizing discussions and progress made, organizing ideas into structured frameworks, synthesizing multiple concepts into cohesive strategies

Feedback and Refinement: Evaluating and improving proposed ideas, offering constructive feedback on suggestions, identifying strengths in proposed approaches, suggesting refinements to enhance messaging effectiveness

Other: Any response that does not clearly fit into the above categories, general pleasantries or acknowledgments, clarification questions about the task, technical assistance unrelated to marketing strategy, administrative comments about the interaction itself

**Response from AI assistant:** {Insert AI Response}

**Output format:** Respond with only the category name. Choose the most appropriate single category from the list above.

## User Evaluation of AI Responses

Critical Evaluation of AI

**Instructions**: You are an expert in analyzing collaborative dialogue. A human user is working with an AI assistant to develop a marketing campaign. You will see the AI assistant's message followed by the user's response. Your task is to classify HOW the user engages with the AI's preceding output.

This is NOT about whether the user's message is long, sophisticated, or well-written. It is about the user's STANCE toward the AI's contribution.

**Classification Categories (1-5):**

1 - **UNCRITICAL ACCEPTANCE:** The user accepts the AI's output without meaningful engagement. They signal agreement, approval, or readiness to move on without evaluating what the AI provided.

Key signals: "sounds good", "yes", "perfect", "I like it", "let's go with that", "great", "ok do it", "looks good to me", thanking and moving on. Also includes: Asking the AI to simply continue, compile, or finalize without any direction change. Delegating the next step without commenting on the current output.

The core test: Could the user have sent this EXACT message regardless of what the AI said? If yes, it is uncritical acceptance.

**2 - PASSIVE DIRECTION:** The user provides a new instruction or preference but does not engage with the substance of what the AI produced. They steer without evaluating. Key signals: "now do the metrics section", "make it more concise", "add something about sustainability", "focus on millennials instead", "can you also include pricing".

The core test: The user is directing the AI but has not indicated whether the AI's previous contribution was good, bad, or needs specific changes. They are moving forward without looking back.

**3 - SELECTIVE ENGAGEMENT:** The user engages with SOME aspect of the AI's output — identifying parts they like or dislike, requesting specific modifications, or accepting some elements while redirecting others. There is evidence they read and processed the AI's response, but engagement is partial.

Key signals: "I like the chef endorsement idea but not the influencer part", "the first two points are strong, can you develop those more?", "good structure but the tone is too corporate", "keep the sustainability angle but rethink the pricing message".

The core test: The user references specific content from the AI's response and makes a judgment about it.

**4 - SUBSTANTIVE EVALUATION:** The user provides reasoned assessment of the AI's output, explaining WHY something works or does not work. They bring their own judgment, domain knowledge, or strategic reasoning to bear on what the AI produced.

Key signals: "The taste-test strategy is smart because it addresses the biggest barrier directly, but celebrity chefs might feel inauthentic for a startup — what about local food critics instead?", "This messaging tries to appeal to both eco-conscious and premium buyers simultaneously, which will dilute both", "The metrics you suggest are all awareness-based but we need conversion metrics to justify the spend".

The core test: The user provides REASONING about the AI's output, not just a preference or direction.

**5 - CRITICAL REFRAMING:** The user challenges the AI's approach, assumptions, or framing. They push back on the direction of the conversation, introduce a substantially different perspective, or identify a fundamental issue with the AI's reasoning.

Key signals: "You're assuming consumers care about sustainability, but premium steak buyers might see lab-grown as downmarket — we need to completely rethink the positioning", "This whole approach is too safe — every lab-grown meat company does education-first, what if we led with exclusivity instead?", "The problem isn't messaging, it's that we're targeting the wrong audience entirely".

The core test: The user is not just modifying the AI's output but challenging or reframing the underlying approach.

Critical Evaluation of AI Cont.

**Important Guidelines:**

- Focus on the user's STANCE toward the AI's output, not message length or writing quality.
- A short message that identifies a specific flaw scores higher than a long message that just says "great, now do the next section."

The key transitions:

- 1→2: User gives direction (vs. just accepting).
- 2→3: User references specific AI content (vs. just directing).
- 3→4: User provides reasoning for their evaluation (vs. just selecting).
- 4→5: User challenges framing or assumptions (vs. just evaluating content).

A user who adds substantial NEW content (their own ideas, domain knowledge, strategic direction) that builds on but goes beyond the AI's output should score at least 3, and higher if they also evaluate what the AI provided.

When uncertain between two adjacent levels, choose the lower one.

**Response Format:**

Respond with ONLY a single integer from 1 to 5. Nothing else.

**AI assistant's message:** {Insert preceding AI response}

**User's response:** {Insert user message}

## Personalization Measurement

Personalization Measurement

**Instructions:** You are evaluating if an AI response is truly personalized to a specific user during a collaborative task (e.g., developing a marketing campaign). Personalization means the AI adapts its content, framing, or interaction style based on the user's unique characteristics, derived from both a synthesized profile of their cognitive, personality, emotional, and creative traits and their professional background/approach.

**General Profile:** {Insert General Profile}

**Interview Summary:** {Insert Interview Summary}

**AI Response:** {Insert AI Response}

**Evaluation Task:** Assess the degree to which the AI Response is genuinely personalized to this specific user, considering all provided user information (synthesized profile + professional summary) and looking for both direct and indirect cues.

**Evaluation Criteria - Consider These Points:**

<u>Explicit Personalization Cues:</u>

Direct Mentions (Synthesized): References to specific personality traits (e.g., "Given your high openness..."), creative strengths (e.g., "Since you excel at generating diverse ideas..."), learning orientation, or other traits from the Synthesized User Profile.
Direct Mentions (Professional): References to specific skills, experiences, job titles, methods, tools, values from the User Persona Summary.
Acknowledging Stated Preferences/Limitations: Adapting to explicitly mentioned factors like lack of experience, collaboration preferences, or specific challenges (often found in the User Persona Summary).

<u>Implicit Personalization Cues:</u>

Tailored Framing/Level: Is the complexity, technicality, or strategic level appropriate for the user\'s described expertise and cognitive capacity?
Relevant Questioning/Suggestions: Are questions/suggestions logically connected to the user\'s professional domain and their cognitive/creative profile? (e.g., Asking a high-openness user for unconventional ideas; suggesting structured steps for a high-conscientiousness user).
Aligned Interaction Style: Does the AI\'s tone and collaborative approach match the user\'s described professional style and personality/emotional traits? (e.g., More enthusiastic for high extraversion; empathetic for high EI; structured for high conscientiousness).
Leveraging Unique Perspective: Does the response implicitly invite contribution based on their unique blend of synthesized traits and professional background?
Indicators of Non-Personalization:
Generic Content: Standard information, templates, or fillers applicable to anyone.
Contradictions: Directly contradicting stated facts or traits in either the synthesized profile or the professional summary.
Ignoring Key Traits: Overlooking central aspects of the user\'s overall profile (synthesized or professional) that should logically influence the interaction.
Irrelevance: Content or questions unrelated to the user\'s background or the task.

**Output Format:**
Provide your analysis in the following JSON format:
{
  "personalized": "Yes/No",
  "type": "Explicit/Implicit/Mixed/NA",
  "elements": ["List of specific user traits addressed in the response, such as: openness, conscientiousness, emotional intelligence, gender, age, experience level, professional background, education, etc."]
}
If the response is not personalized, use "NA" for type and an empty list for elements.

| Trait | Trait-Related Question |
| --- | --- |
| Openness to Experience | If I wanted to take this Cultivated ribeye campaign in a direction that breaks conventional food marketing norms, what unexplored creative territory would be most worth my time to investigate? |
| Conscientiousness | To ensure this campaign development process is thorough and effective from the start, what foundational step or organizational framework should I prioritize setting up before we get into the creative details? |
| Extraversion | When I need to develop breakthrough ideas for this ribeye campaign, should I schedule a team brainstorming session or block off some solo thinking time first? |
| Agreeableness | If a key team member raises a reasonable objection to my campaign idea, what should my primary focus be during that conversation to handle it constructively and maintain a positive working dynamic? |
| Neuroticism | Considering the risks and potential public skepticism around this product, how should I frame my approach mentally to stay effective and focused, even with setbacks or criticism? |
| Emotional Affect | Facing the marketing challenges for cultivated meat, what's the most uplifting perspective or exciting potential I should focus on to stay motivated? |
| Emotional Intelligence | How should I approach understanding what consumers might be feeling about cultivated meat before I craft my messaging? |
| Learning Goal Orientation | How can I set up my campaign process so my strategy evolves based on ongoing learning about what's working with consumers and what isn't? |
| Fluid Intelligence | When explaining the science behind Cultivated, what's the right level of detail I should aim for in my marketing to be clear without being overly technical for the audience? |
| Divergent Thinking | If I wanted to create an unexpected marketing approach for Cultivated ribeye, what direction would be worth exploring that hasn't been done before in food marketing? |
| Problem-Solving | To choose the best call-to-action ('Learn More', 'Find Retailers', 'Reserve Yours') for my initial ads, what's the very first action or analysis I should undertake, staying consistent with how I usually approach these marketing decisions? |
| Core Values | As I craft the core explanation of what Cultivated meat is, what fundamental element reflecting my typical work focus should be the cornerstone of my message? |
| Collaboration Style | After I've drafted this tagline ('Cultivated Ribeye: Peak Steak, Peak Sustainability'), what would be my go-to approach for validating it before finalizing? |
| Risk Tolerance | Regarding the campaign launch strategy, should I lean towards a proven, safer approach with predictable results, or explore a more innovative, potentially higher-impact tactic that carries more risk? |
| Adaptability | Given a very tight budget for this initial launch phase, how should that constraint influence my choice between focusing on broad awareness versus targeting a very specific niche? |
| Marketing Background | Given Cultivated's novelty and potential skepticism, where should the majority of our initial marketing budget (first 3-6 months) be allocated for maximum long-term success: heavily towards direct product trial and purchase activation (e.g., pop-up tasting events, introductory offers, easy online reservations) OR heavily towards foundational education and credibility building (e.g., detailed process explainers, securing top chef endorsements, third-party safety validations) before a wider sales push? |

**Table S1. Open-ended questions to elicit personalized AI responses along a single dimension.**

# Results

## Personalization of AI Responses

On average, each AI response contains 6.1 (SD: 1.2) personalized elements out of 22 possible traits mentioned across all responses indicating a multifaceted approach to personalization. Among those responses, conscientiousness (78.6%), marketing experience (66.6%), and openness to experience (65.9%) are the traits most referenced by the assistant (**Fig. S4B**). Notably, there is no increase in the proportion of personalization over time in the generic AI condition, indicating that the system did not appreciably 'learn' more about its users. Lastly, we checked to see how different individual profiles were from one another among participants in the partial personalization condition (**Fig. S4D**). Overall, complete profiles (psychological + marketing experience) tend to be very similar ($\text{Cosine Similarity}_{\text{Mean}}$ = 0.88, SD = 0.03) with greater variance among the marketing experience summary ($\text{Cosine Similarity}_{\text{Mean}}$ = 0.69, SD = 0.08) than the psychological component ($\text{Cosine Similarity}_{\text{Mean}}$ = 0.89, SD = 0.04). Given that the interview was semi-structured, driven in large part by what the user chose to disclose, it is unsurprising that there is larger intra-individual variability between true and randomly assigned interview summaries than between psychological profiles.

Although certain characteristics may be frequently used to tailor a system's response, frequency alone does not tell us about a trait's relative importance, e.g., is it more important to personalize on marketing experience or agreeableness? Too much irrelevant context may be detrimental to model performance, potentially undermining the benefits of personalization. To systematically evaluate which traits offer the greatest personalization potential, we conducted a simulation study comparing AI responses (between personalized vs. generic prompts) across 16 dimensions. We find clear differences in personalization effectiveness across traits. Learning goal orientation, problem solving, and core values produce the largest differentiation between personalized and generic responses, with cosine similarity values ranging from approximately 0.7-0.8 (**Fig. S5**). However, these traits are not frequently referenced by the AI assistant. This discrepancy demonstrates that AI's naturalistic approach to personalization may not optimally utilize the entirety of an individual's relevant data. Conversely, traits related to emotional processing (emotional intelligence, affect, and agreeableness) show smaller differences between AI responses, suggesting that these traits may be less critical for effective personalization within a work relevant task.

## Results are Robust to Other LLMs and Prompts

As a sensitivity analysis, we generated campaign ratings for i) average of 4 smaller capability LLMs (Llama 3.1, Mistral Small, Claude 3.5 Haiku, and Gemini 1.5 Flash) and ii) 3 larger capability LLMs (GPT-4o, Gemini 2.5 Flash, and Claude 4 Sonnet) and re-ran our regression model (**Table S3**). The personalized treatment condition leads to significantly better campaigns when evaluated by both smaller ($\beta = 0.35$, SE = 0.16, $p = 0.03$) and larger ($\beta = 0.36$, SE = 0.18, $p = 0.05$) LLMs, thereby showing that our results are not dependent on the choice of LLM. To ensure that our findings are only because of personalization, we tested 3 other campaign assessment prompts: i) a different company but same evaluation criteria, ii) entirely unrelated task (employee performance reviews), and iii) alternative evaluation framework. As expected, there is no significant personalization effect when campaigns are evaluated with prompts designed for either a different company ($\beta = 0.09$, SE = 0.14, $p = 0.53$) or unrelated task ($\beta = -0.04$, SE = 0.06, $p = 0.53$) (**Table S4**). However, personalization is still significantly associated with better marketing campaigns when using an alternative set of evaluation criteria (β = 0.50, SE = 0.19, p = 0.008). We can therefore conclude that personalized AI enhances a person's ability to generate high quality content and that this effect is specific to the personalization of human-AI interactions – not because of differences in either LLM ratings or an inability of the evaluation criteria to assess domain-specific quality attributes.

## Elevated Failures of Conversational Grounding in Generic AI

Communication between humans-AI can occasionally become sidetracked due to a variety of factors, e.g., misunderstandings, expectation misalignment, etc. We examined the effect of personalization on the relative frequency of corrective acts, i.e., turns with insufficient conversational grounding. AI initiated substantially fewer corrective acts to address grounding than human users (**Fig. S6**), in line with previous work [7]. A higher proportion of corrective acts indicates a failure in communication, potentially undermining collaboration. For example, in the personalized AI condition human users attempt to address a lack of adequate conversational grounding in 8.9% of turns, whereas the corresponding AI assistant responded with a corrective act in only 1.5% of its turns. Among the three AI assistants, generic AI addressed failures of conversational grounding more frequently than either personalized (β = 0.03, SE = 0.01, p = 0.04) or partially personalized AI (β = 0.04, SE = 0.01, p = 0.01), driven by a propensity to ask more clarification questions. This may also be due in part to users signaling more grounding failures when interacting with generic AI.

## Personalization Shifted Conversations from Collaborative to Strategic

We quantified language changes – in greater depth than act and coordinate turn classification – focusing on brainstorming (human), content development (AI), delegation to AI (human), and strategic guidance (AI) that together account for core features of AI utilization. Among these 4 categories, we found that partially personalized AI caused the largest effect, relative to generic AI (**Fig. S8**). Conversations with partially personalized AI conveyed significantly more content development (β = 0.09, SE = 0.03, p = 0.003), delegation acts (β = 0.11, SE = 0.04, p = 0.006), and strategic guidance questions (β = 0.13, SE = 0.04, p = 0.002). Because generic AI users received less complete responses, and more collaborative type follow-up questions, they edited their final AI-suggested marketing campaigns substantially more prior to submission. Approximately 80.5% of words were different between the final campaign and what their AI assistant recommended versus 64.5% in the personalized arm. Consequently, the increased time participants spent reworking the campaign's core text reduced the time available for crucial strategic components, such as creativity and feasibility.

Lastly, we examined differences in how participants prompted the AI assistant as another, independent, explanation for variation in performance. We embedded the combined user turns - at the individual level - as text vectors, reduced their dimensionality using UMAP, and then identified 3 distinct clusters (**Fig. S9**). Broadly the three clusters can be described as “collaborative iteration” (Cluster 1), “structured professional specifications” (Cluster 2), “strategic direction and execution” (Cluster 3). Cluster 3 type messages were associated with significantly better marketing campaigns than both Cluster 1 ($\beta = 0.40$, SE = 0.19, $p = 0.03$) and Cluster 2 ($\beta = 1.0$, SE = 0.21, $p < 0.001$). Additionally, the association between message cluster and performance remained significant after controlling for the number of marketing skills. From these different analyses, strategic guidance emerged as an important feature driving the development of better marketing campaigns. By supporting users’ attention, of which strategic guidance is one component, and reasoning personalized AI facilitates more directed creative collaboration than is otherwise possible with a generic assistant.

| | Generic (n = 116) | Personalized (n = 114) | Partial Personalization (n = 101) | p-value |
|---|---|---|---|---|
| **Age (years)** | 38.7 (12.7) | 38.6 (11.8) | 38.8 (11.3) | 0.99 |
| **Gender** | | | | |
| Female | 63 (54.3%) | 52 (45.6 %) | 50 (49.5%) | 0.56[a] |
| Male | 52 (44.8%) | 61 (53.5%) | 50 (49.5%) | - |
| Other | 1 (0.9%) | 1 (0.9%) | 1 (1.0%) | - |
| **Education** | | | | |
| High school | 16 (13.8%) | 9 (7.9%) | 7 (6.9%) | 0.04*[a] |
| Some university | 17 (14.7%) | 13 (11.4%) | 7 (6.9%) | - |
| Associate degree | 6 (5.2%) | 9 (7.9%) | 5 (5.0%) | - |
| Bachelor's degree | 50 (43.1%) | 55 (48.2%) | 39 (38.6%) | - |
| Master's degree | 24 (20.7%) | 24 (21.1%) | 33 (32.7%) | - |
| Professional degree | 0 (0.0%) | 1 (0.9%) | 0 (0.0%) | - |
| Doctorate | 3 (2.6%) | 3 (2.6%) | 10 (9.9%) | - |
| **Currently employed (% Yes)** | 109 (94.0%) | 109 (95.6%) | 95 (94.1%) | 0.37[a] |
| **Expertise type** | | | | |
| Marketing | 34 (29.3%) | 37 (32.5%) | 32 (31.7%) | 0.86[a] |
| Product | 37 (31.9%) | 40 (35.1%) | 36 (35.6%) | - |
| Neither | 45 (38.8%) | 37 (32.5%) | 33 (32.7%) | - |
| **Number of marketing skills (0-21)** | 7.7 (1.5) | 7.4 (1.3) | 7.5 (1.5) | 0.19 |
| **Ever used an LLM (% Yes)** | 111 (95.7%) | 109 (95.6%) | 98 (97.0%) | 0.84[a] |
| **LLM usage past week (Days)** | 3.8 (2.2) | 3.8 (2.0) | 3.9 (2.1) | 0.96 |

**Table S2. Participant Demographic Characteristics.** [a]Chi-Squared Test

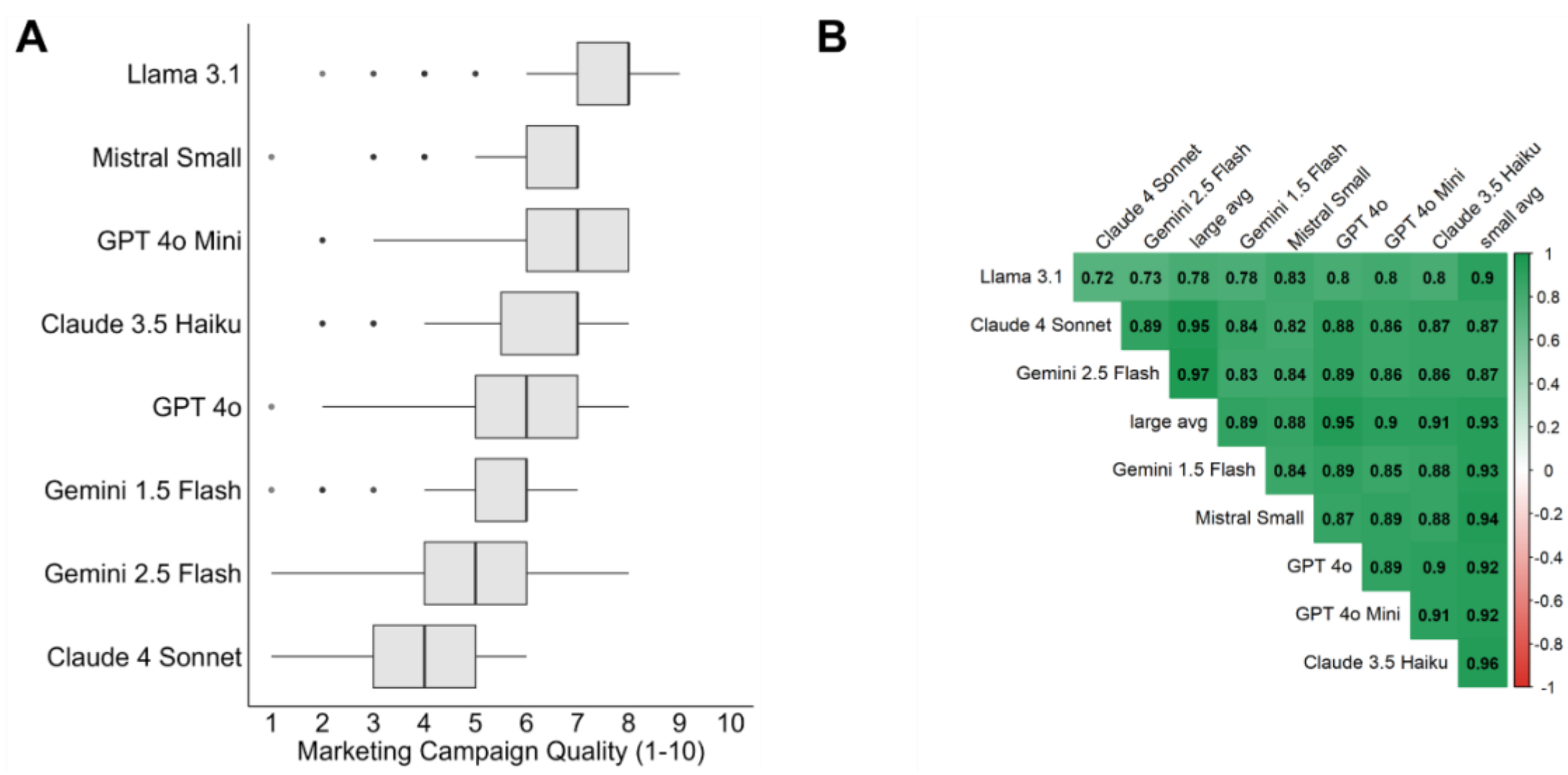

**Fig. S1. Marketing Campaign Quality Ratings for Different LLMs.**

**(A)** Absolute Marketing Campaign Quality Ratings for 8 LLMS, 3 large capacity models (Claude 4 Sonnet, Gemini 2.5 Flash, GPT 4o) and 5 smaller models (Llama 3.1, Mistral Small, GPT 4o-mini, Claude 3.5 Haiku, and Gemini 1.5 Flash). Small models tended to rate the campaigns more favorably, median >6/10, than larger ones. **(B)** Despite differences in absolute ratings, all LLM-as-a-research assistant evaluations were strongly correlated with one another ($r > 0.72$, $p < 0.001$).

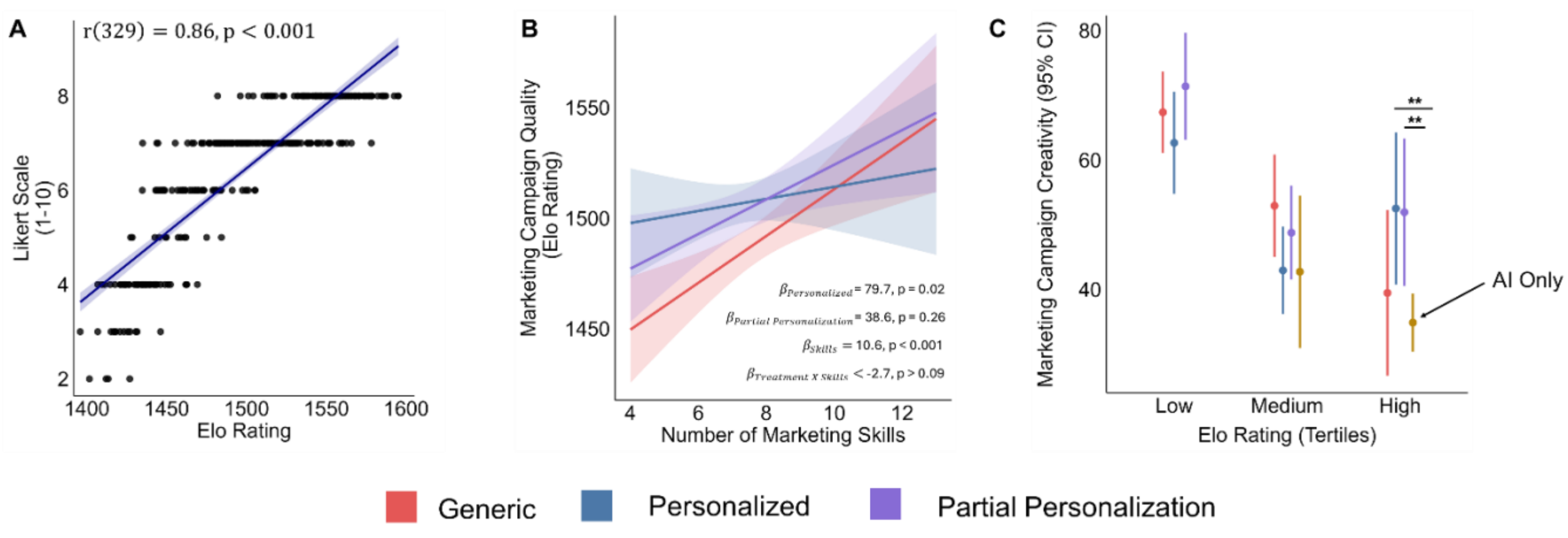

**Fig. S2. LLM Campaign Quality Elo Ratings.**

**(A)** Significant positive association between LLM Elo ratings and Likert scale ratings ($r(329) = 0.86$, $p < 0.001$). **(B)** Interaction between AI treatment (generic, personalized, and partial personalization) and marketing skills, after controlling for number of turns with the AI assistant. There is a significant main effect of personalized AI ($\beta = 79.7$, $SE = 35.4$, $p = 0.02$) and the number of marketing-relevant skills (β = 10.6, SE = 3.1, $p < 0.001$), but not partially personalized AI (β = 38.6, SE = 34.4, $p = 0.26$). There is no significant interaction between personalized AI and number of skills (β = -7.9, SE = 4.6, $p = 0.09$). **(C)** Among high quality campaigns (top third), personalized (β = 17.6, SE = 5.5, $p = 0.002$) and partially personalized (β = 17.1, SE = 6.5, $p = 0.01$) AI significantly increases campaign creativity relative to the generic treatment arm. Although approximately 80% of AI only campaigns are of high quality, they are consistently less creative than those created through human-AI collaboration.

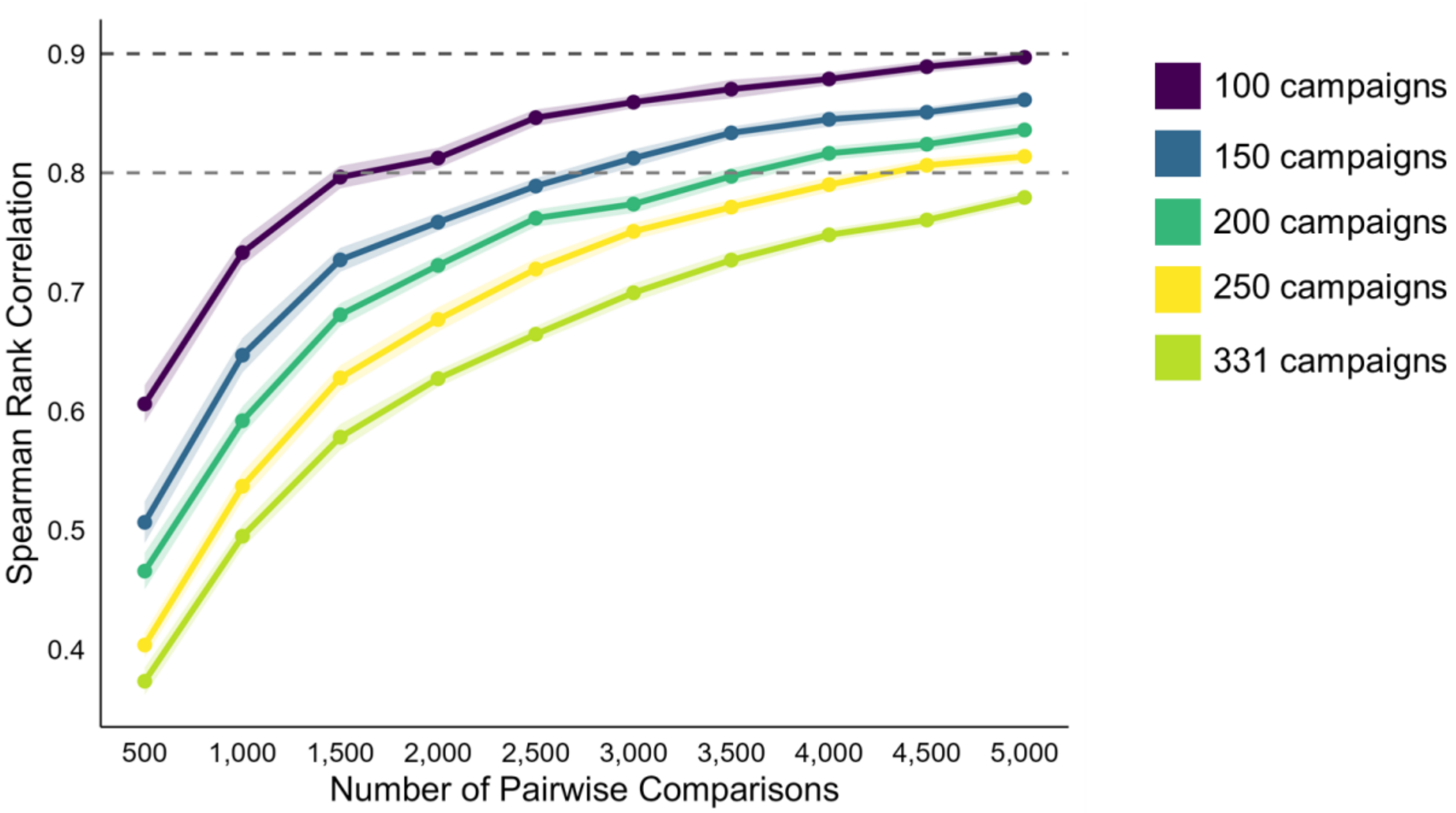

**Fig. S3. Power Simulation to Identify Number of Campaigns Ratings Required to Achieve Adequate Elo Ranking Stability (ρ > 0.8).**

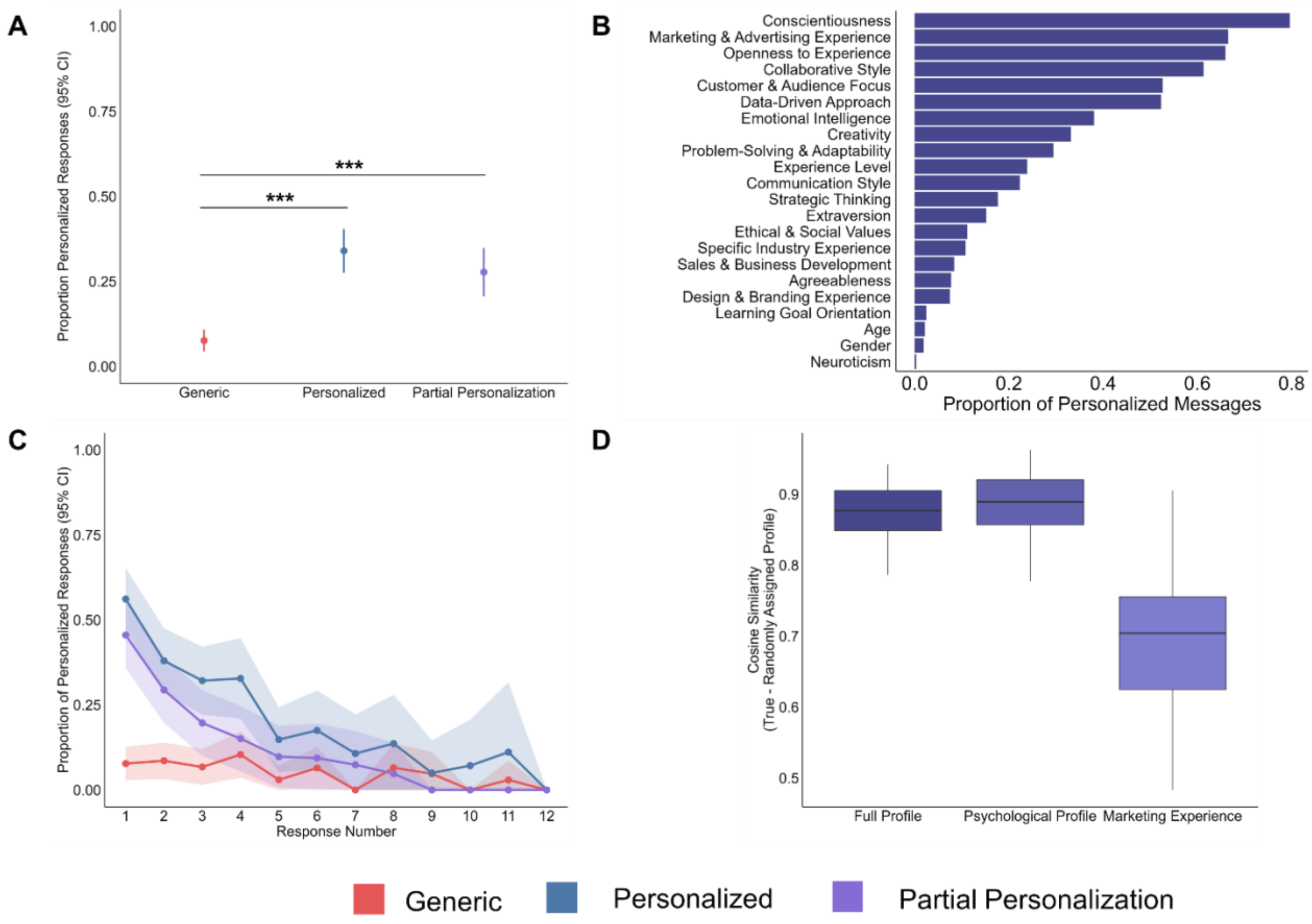

**Fig. S4. Personalization of AI Turns**

**(A)** Proportion of personalized responses by treatment condition. Compared to generic AI, personalized ($\beta = 0.26$, SE = 0.04, $p < 0.001$) and partially personalized AI ($\beta = 0.20$, SE = 0.04, $p < 0.001$) had significantly higher rates of personalized responses. **(B)** Proportion of various traits referenced within personalized responses. Conscientiousness, marketing experience, and openness to new experience were the most frequently referenced traits. Age, gender, neuroticism, meanwhile, were the least commonly used personalized elements, found in 2% or less of responses. **(C)** Proportion of personalized responses by conversational round (response number). **(D)** Cosine similarity of true vs. randomly assigned profile among participants in the partial personalization group for: i) full profile, ii) psychological profile only, and iii) marketing experience profile.

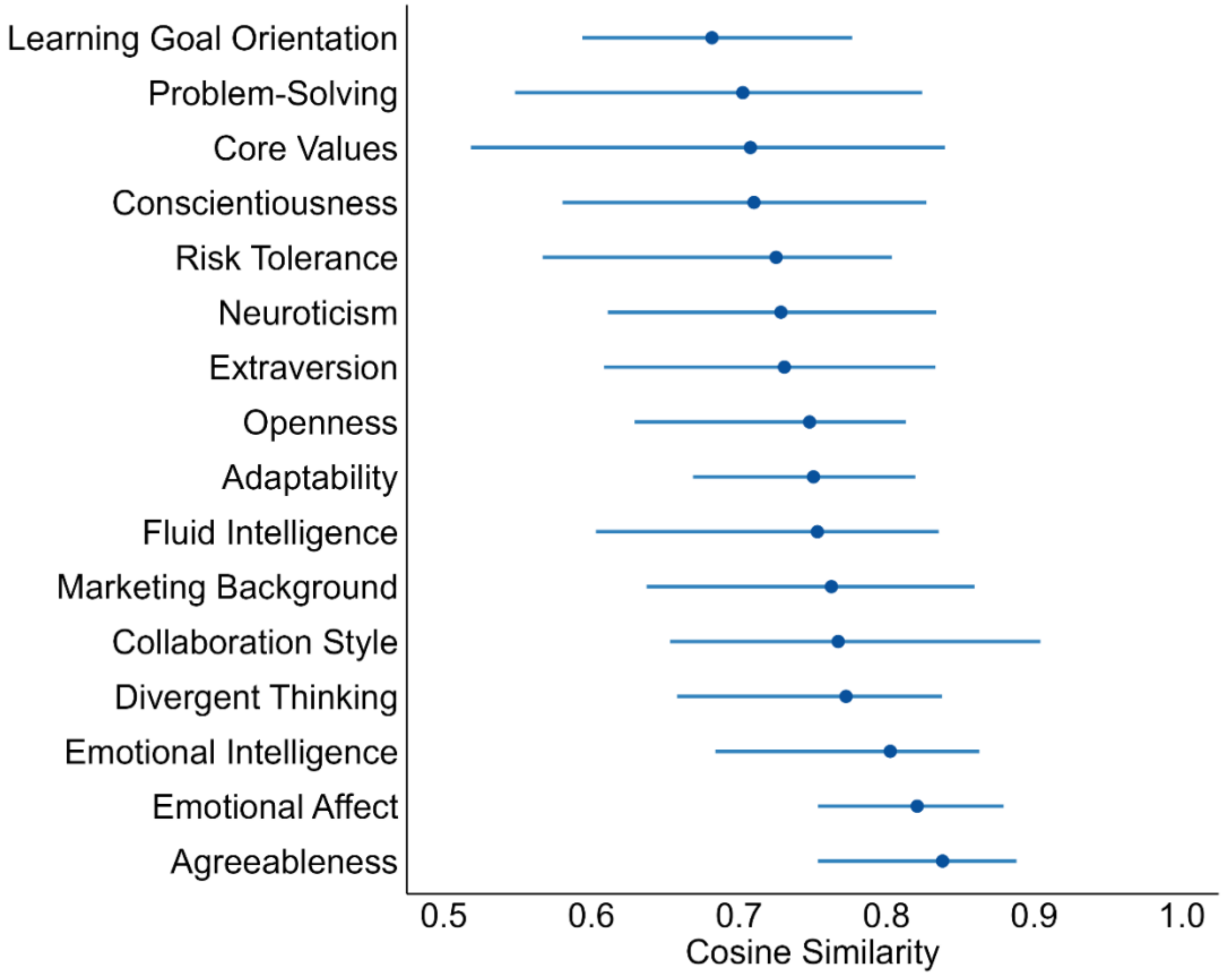

**Fig. S5. Difference between AI Responses (Personalized vs. Generic) from Questions Designed to Elicit a Personalized Response on One Attribute**

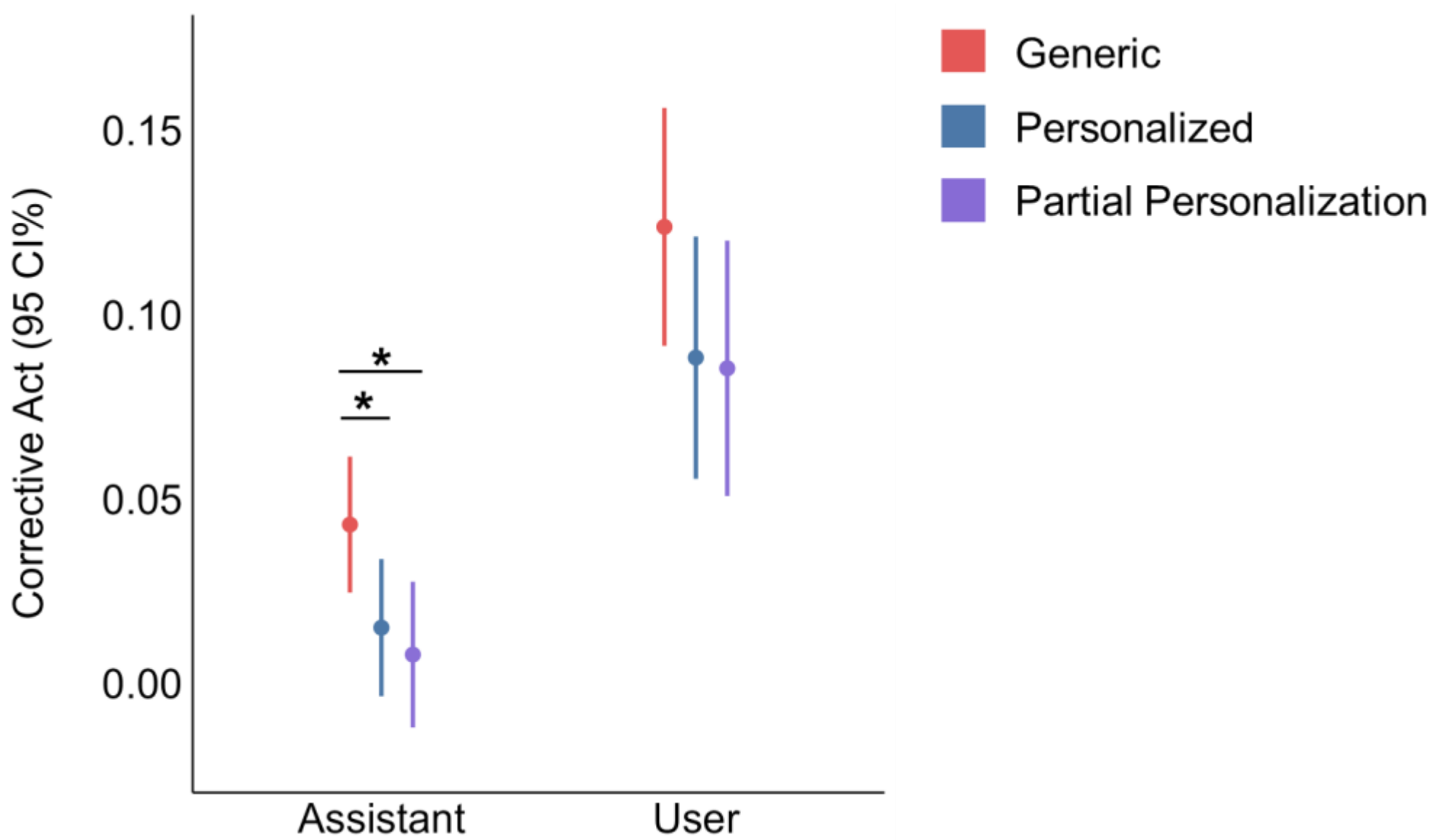

**Fig. S6. Proportion of Corrective Acts by AI Assistant and Human User.**

Personalized (β = -0.04, SE = 0.02, p = 0.13) and partially personalized AI (β = -0.04, SE = 0.02, p = 0.11) have significantly fewer corrective acts, i.e., clarification, topic switch, repair, repeat, and disagreement compared to the generic assistant. Although human users send substantially more (~2x) corrective acts in general, there is no significant difference between generic and personalized treatments.

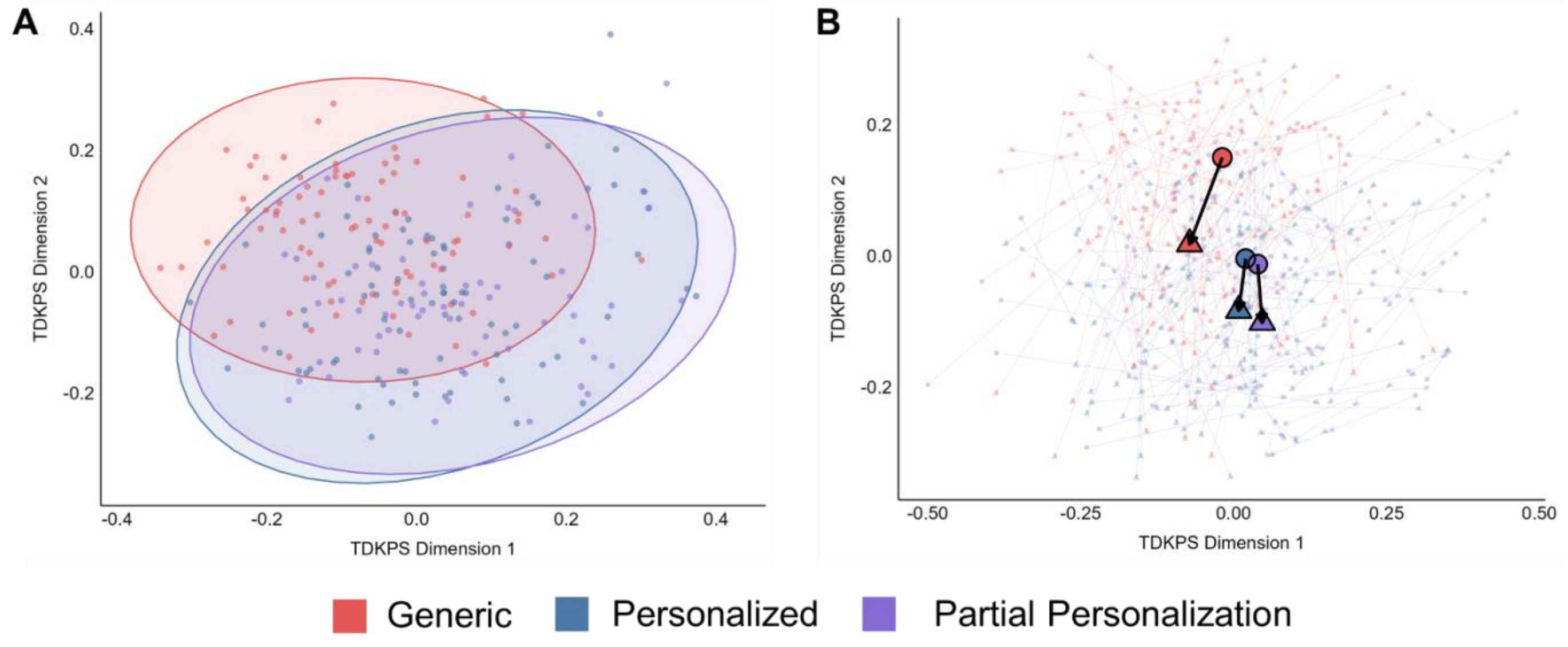

**Fig. S7. Distributional Shift Analysis of AI Response Embeddings.**

**(A)** Two-dimensional CMDS projection of participant-averaged AI response embeddings. The generic condition is clearly separated from both personalized and partially personalized conditions ($p < 0.001$), while the two personalized conditions overlap substantially ($p = 0.49$). **(B)** Temporal trajectories showing how AI response distributions shift from the first to second half of the conversation. Arrows indicate participant-level shifts; large markers show condition means. While all conditions shifted in broadly similar directions (pairwise cosine similarities: 0.72–0.80), generic AI drifted along a significantly different trajectory than both the personalized ($p = 0.005$) and partially personalized ($p = 0.002$) conditions, while the two personalized conditions did not drift apart significantly ($p = 0.34$)

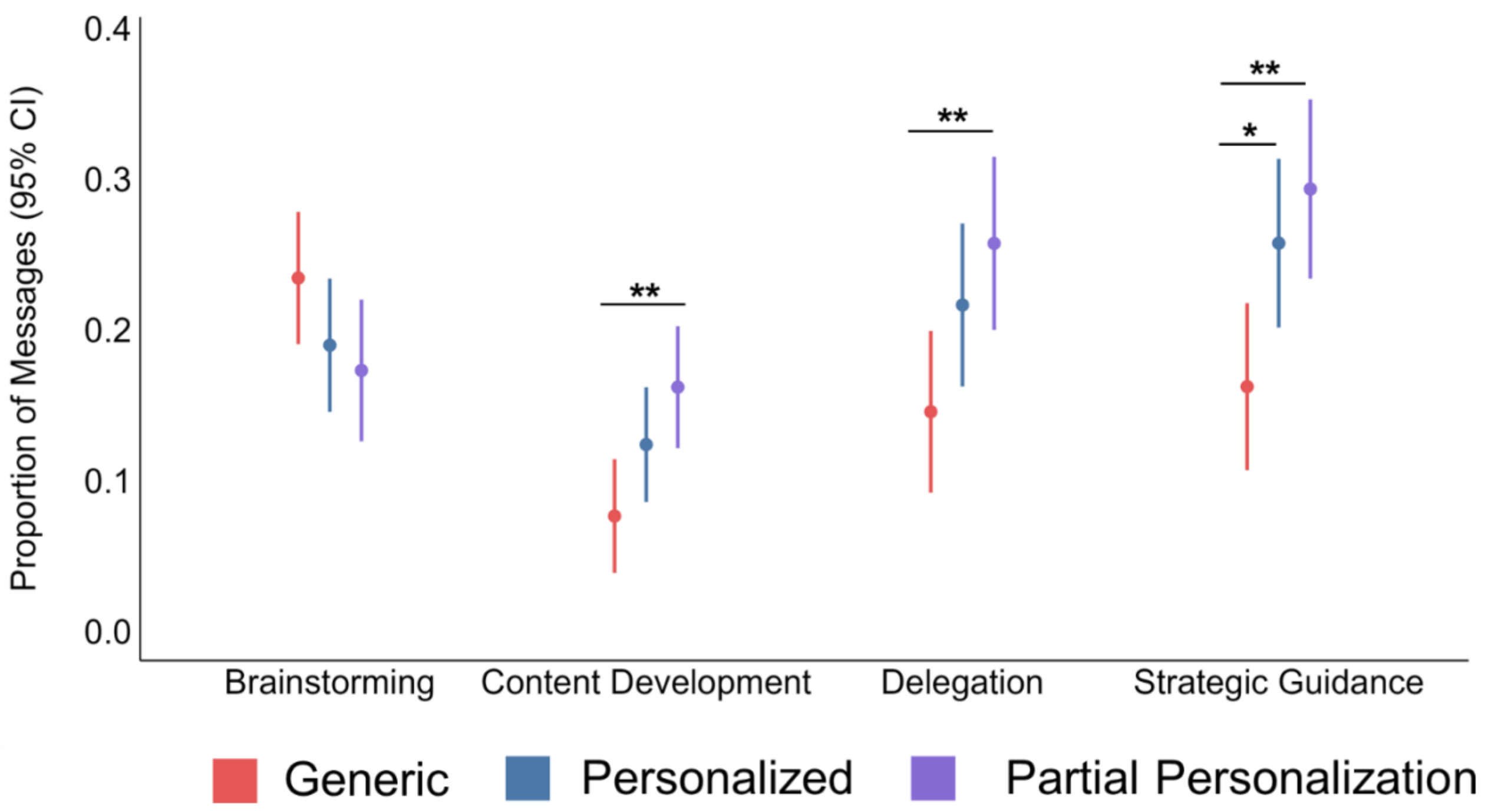

**Fig. S8. Turn Type in Human-AI Conversational Dialogue.**

Participants working with partially personalized AI receive significantly more content development ($\beta = 0.09$, $SE = 0.03$, $p = 0.003$) and strategic guidance ($\beta = 0.13$, $SE = 0.04$, $p = 0.002$) responses, and delegate ($\beta = 0.11$, $SE = 0.04$, $p = 0.006$) significantly more to their AI assistant. There is no significant difference among AI assistants regarding the frequency of brainstorming-related turns.

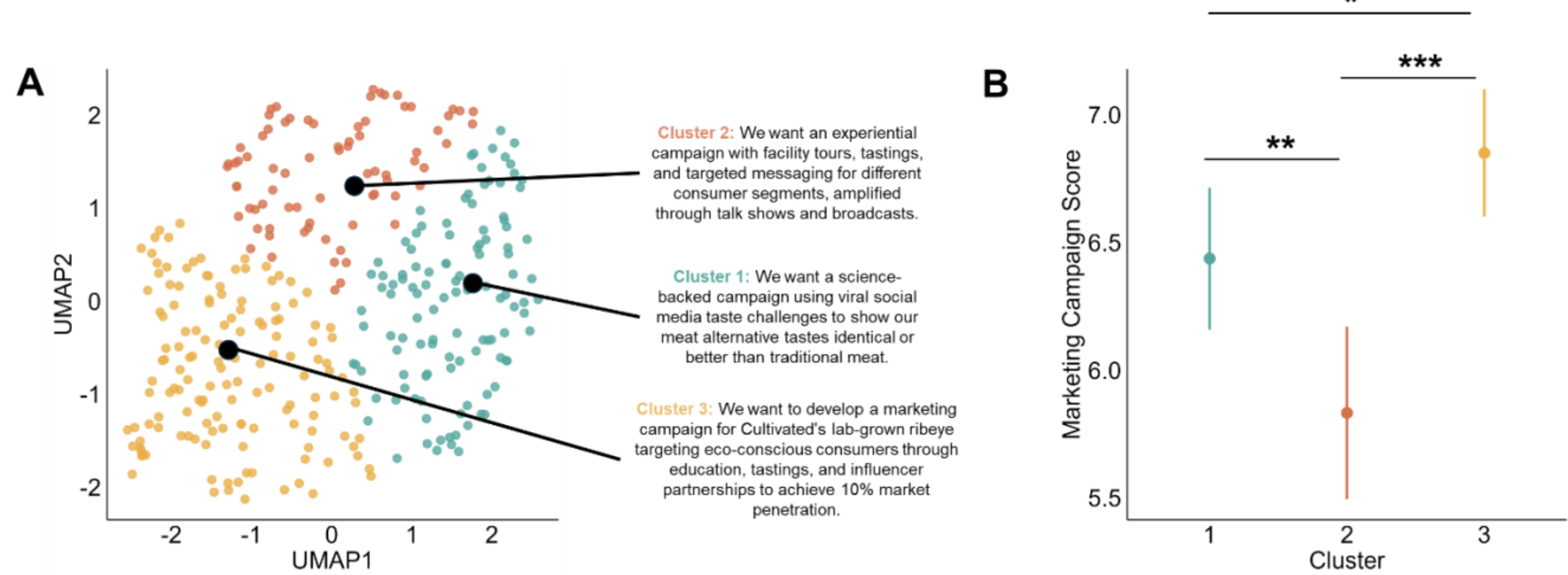

**Fig. S9. Classification of Human Messages.**

**(A)** Across treatment conditions, human messages are found to cluster into 3 groups: collaborative ideation (Cluster 1), professional specifications (Cluster 2), and strategic guidance (Cluster 3). **(B)** Participants who send messages with a strong emphasis on strategic guidance, i.e., Cluster 3, produced significantly better campaigns than either Cluster 1 ($\beta = 0.40$, SE = 0.19, $p = 0.03$) or Cluster 2 ($\beta = 1.0$, SE = 0.21, $p < 0.001$).

| Reference (Generic) | **GPT 4o-mini** | **Avg. Small LLMs** | **Avg. Large LLMs** |
|---|---|---|---|
| Personalized | 0.59 (0.21)** | 0.35 (0.16)* | 0.36 (0.18)* |
| Partial Personalization | 0.56 (0.21)** | 0.34 (0.16)* | 0.26 (0.17) |
| Turn Count | 0.06 (0.02)** | 0.05 (0.01)** | 0.03 (0.02) |

**Table S3. Personalized AI is Associated with Higher Marketing Campaign Quality for LLM-as-a-Judge Evaluations Regardless of Model Size**

| | **(1) Simplified Evaluation Criteria** | **(2) Different Company** | **(3) Entirely Unrelated Task** | **(4) Modified Evaluation Criteria** |
|---|---|---|---|---|
| Personalized | 0.10 (0.07) | 0.09 (0.14) | -0.04 (0.06) | 0.50 (0.19)** |
| Partial Personalization | 0.10 (0.07) | 0.01 (0.14) | -0.08 (0.06) | 0.44 (0.19)* |
| Turn Count | 0.02 (0.006)** | 0.02 (0.005) | -0.01 (0.005) | 0.07 (0.02)*** |

**Table S4. Sensitivity Analysis for Different Types of Campaign Evaluation Prompts.**

(1) Same prompt used by human raters to judge overall campaign quality on a Likert scale of 1-5 (2) An entirely different company, i.e., B2B cybersecurity startup, but same evaluation framework. (3) A task that is irrelevant to the main experiment, i.e., employee performance review, with a completely different set of evaluation criteria (4) Alternative evaluation criteria compared to what was used in the main experimental analyses. There was a significant effect of treatment on campaign quality only for the modified evaluation criteria prompt, indicating the main effects are robust to qualitatively different marketing campaign evaluations. But there is no effect of personalization when using a simplified rubric (1) or on tasks unrelated to the development of a marketing campaign (2 & 3).

**Confirmatory Factor Analysis**

We measure latent construct of collective memory, attention, and reasoning through several behavioral and self-reported survey measures (**Table S5**). All indicators load significantly on their intended factors (β = 0.48 – 0.9, $p < 0.001$). Internal consistency of all factors is acceptable exceeding 0.68 for all factors (Memory: α = 0.68 [95% CI: 0.61 – 0.74]; Attention: α = 0.70 [95% CI: 0.64 – 0.74]; Reasoning: α = 0.87, [95% CI: 0.85-0.89]). Latent correlations range from -.38 to .22, all below .85, supporting discriminant validity. A confirmatory factor analysis of the hypothesized 3-factor model indicates acceptable to good fit of the latent structure ($\chi^2(51) = 97.072$, $p < .001$; CFI = .968; TLI = .959; RMSEA = .052, 90% CI [.036, .068]; SRMR = .05).

| Item | Measurement Type | Standardized Loading | p-value |
|---|---|---|---|
| **Factor 1: Memory** | | | |
| Campaign Word Count | Behavioral | 0.514 | <0.001 |
| Campaign Components | LLM assessed | 0.607 | <0.001 |
| Marketing Terms Word Count | LLM assessed | 0.864 | <0.001 |
| **Factor 2: Attention** | | | |
| Collaborative Ideation | LLM assessed | 0.661 | <0.001 |
| AI Response | Behavioral (Flesch) | 0.699 | <0.001 |
| Strategic Guidance (reverse coded) | LLM assessed | 0.484 | <0.001 |
| Delegation (reverse coded) | LLM assessed | 0.571 | <0.001 |
| **Factor 3: Reasoning** | | | |
| Task Feedback | Self-report | 0.901 | <0.001 |
| Task Assistance | Self-report | 0.881 | <0.001 |
| Recommendation | Self-report | 0.875 | <0.001 |
| Cognitive Support Types | Self-report | 0.526 | <0.001 |
| Confidence in AI | Self-report | 0.625 | <0.001 |

**Table S5. Collective Intelligence Confirmatory Factor Analysis for 3-Factor Structure (Memory, Attention, and Reasoning)**

## References

1. Shi, F., et al. *Large language models can be easily distracted by irrelevant context*. in *International Conference on Machine Learning*. 2023. PMLR.
2. Liusie, A., P. Manakul, and M. Gales. *LLM comparative assessment: Zero-shot NLG evaluation through pairwise comparisons using large language models*. in *Proceedings of the 18th conference of the European chapter of the Association for Computational Linguistics (volume 1: long papers)*. 2024.
3. Zhang, D.-W., et al., *Evaluating large language models for criterion-based grading from agreement to consistency.* npj Science of Learning, 2024. **9**(1): p. 79.
4. Wu, S., et al., *Aligning llms with individual preferences via interaction.* arXiv preprint arXiv:2410.03642, 2024.

5. Pham, C.M., et al., *Topicgpt: A prompt-based topic modeling framework. DOI: 10.48550.* arXiv preprint arXiv.2311.01449, 2024.
6. Bridgeford, E. and H. Helm, *Detecting Perspective Shifts in Multi-agent Systems.* arXiv preprint arXiv:2512.05013, 2025.
7. Shaikh, O., et al., *Navigating Rifts in Human-LLM Grounding: Study and Benchmark.* arXiv preprint arXiv:2503.13975, 2025.